\documentclass[11pt]{article}

\usepackage[final]{acl}

\usepackage{times}
\usepackage{latexsym}
\usepackage{booktabs}
\usepackage{multirow}
\usepackage{multicol}
\usepackage{amssymb}
\usepackage{pifont}
\usepackage{xcolor}
\usepackage{cleveref}
\usepackage{subcaption}
\usepackage{longtable}
\usepackage{fancybox}

\usepackage{adjustbox}
\usepackage{booktabs}
\usepackage{multirow}
\usepackage{pifont}

\let\oldding\ding
\renewcommand{\ding}[1]{%
  \ifnum#1=52
    \textcolor{green}{\oldding{#1}}%
  \else
    \ifnum#1=56
      \textcolor{red}{\oldding{#1}}%
    \else
      \ifnum#1=51
        \textcolor{orange}{\oldding{#1}}
      \else
      \fi
    \fi
  \fi
}

\definecolor{LMUGrey3}{HTML}{626468}
\usepackage[framemethod=tikz]{mdframed}
\newmdenv[
  backgroundcolor=LMUGrey3!10,
  linecolor=LMUGrey3!10,
  innertopmargin=2pt,
  innerbottommargin=2pt,
  innerleftmargin=2pt,
  innerrightmargin=2pt,
  skipabove=10pt,
  skipbelow=10pt,
  leftmargin=0pt,
  rightmargin=0pt,
  linewidth=0pt,
]{lightbox}

\usepackage{acronym}
\acrodef{LLM}{large language model}
\acrodef{ALM}{Audio Language Model}
\acrodef{LALM}{large audio language model}
\acrodef{LAM}{Large Audio Model}
\acrodef{ARM}{Audio Reasoning Model}
\acrodef{LARM}{Large Audio Reasoning Model}
\acrodef{VLM}{Vision Language Model}
\acrodef{LVLM}{Large Vision Language Model}
\acrodef{MLLM}{multimodal large language model}
\acrodef{MCQ}{multiple choice question}

\usepackage[T1]{fontenc}

\usepackage[utf8]{inputenc}

\usepackage{microtype}

\usepackage{inconsolata}

\usepackage{graphicx}

\title{EXAM$^2$: \underline{Ex}tending \underline{A}udio Understanding in \underline{M}ultilingual and \underline{M}ultimodal Analysis}

\author{
Jiawen Wang$^{1}$ \quad Xiaoxue Gao$^{2 \dagger}$ \quad Zi Haur Pang$^{3}$ \quad Nancy F. Chen$^{4}$ \\
\\
$^1$University of Munich, Germany\\
$^2$School of Artificial Intelligence, The Chinese University of Hong Kong, Shenzhen, China\\
$^3$Kyoto University, Japan \enspace $^4$A*STAR, Singapore\\
\texttt{jiawen.wang@campus.lmu.de} 
}

\begin{document}
\maketitle

\begingroup
\renewcommand{\thefootnote}{}
\footnotetext{\textdagger\ Corresponding author.} 
\endgroup

\begin{abstract}
Recent large audio language models (LALMs) have achieved impressive progress in audio understanding. 
However, existing evaluations remain largely constrained to English and narrow audio domains. 
Prior benchmarks typically focus on a single audio modality, i.e., speech, sound, or music, limiting the systematic investigation into how these models generalize across diverse visual scenarios. 
In this paper, we introduce EXAM$^2$, a benchmark for multilingual and multimodal audio understanding spanning six languages and multiple modalities, including speech, sound, music, mixed-audio settings, and visual images.
By incorporating visual information alongside heterogeneous audio inputs, EXAM$^2$ enables more realistic evaluation of scene-aware audio reasoning and cross-modal comprehension. 
EXAM$^2$ comprises $5,667$ multiple-choice questions, $22,614$ image instances, and $135,684$ multilingual translations. 
We evaluate state-of-the-art open-source and proprietary LALMs as well as multimodal LLMs, revealing substantial performance gaps in multilingual and cross-modal understanding. 
Furthermore, we propose Gemma3n-EXAM$^2$, a lightweight fusion-model fine-tuned on EXAM$^2$-train, achieves up to $15.8\%$ improvement in multilingual settings and $16.5\%$ gains in multimodal evaluation over a strong baseline. 
Empirical results establish EXAM$^2$ as a challenging benchmark and pioneer future multilingual and multimodal audio intelligence research.\footnote{All code, data, and models are available on \url{https://github.com/werywjw/EXAM-2}.}
\end{abstract}

\begin{figure*}[ht!]
  \includegraphics[width=\linewidth]{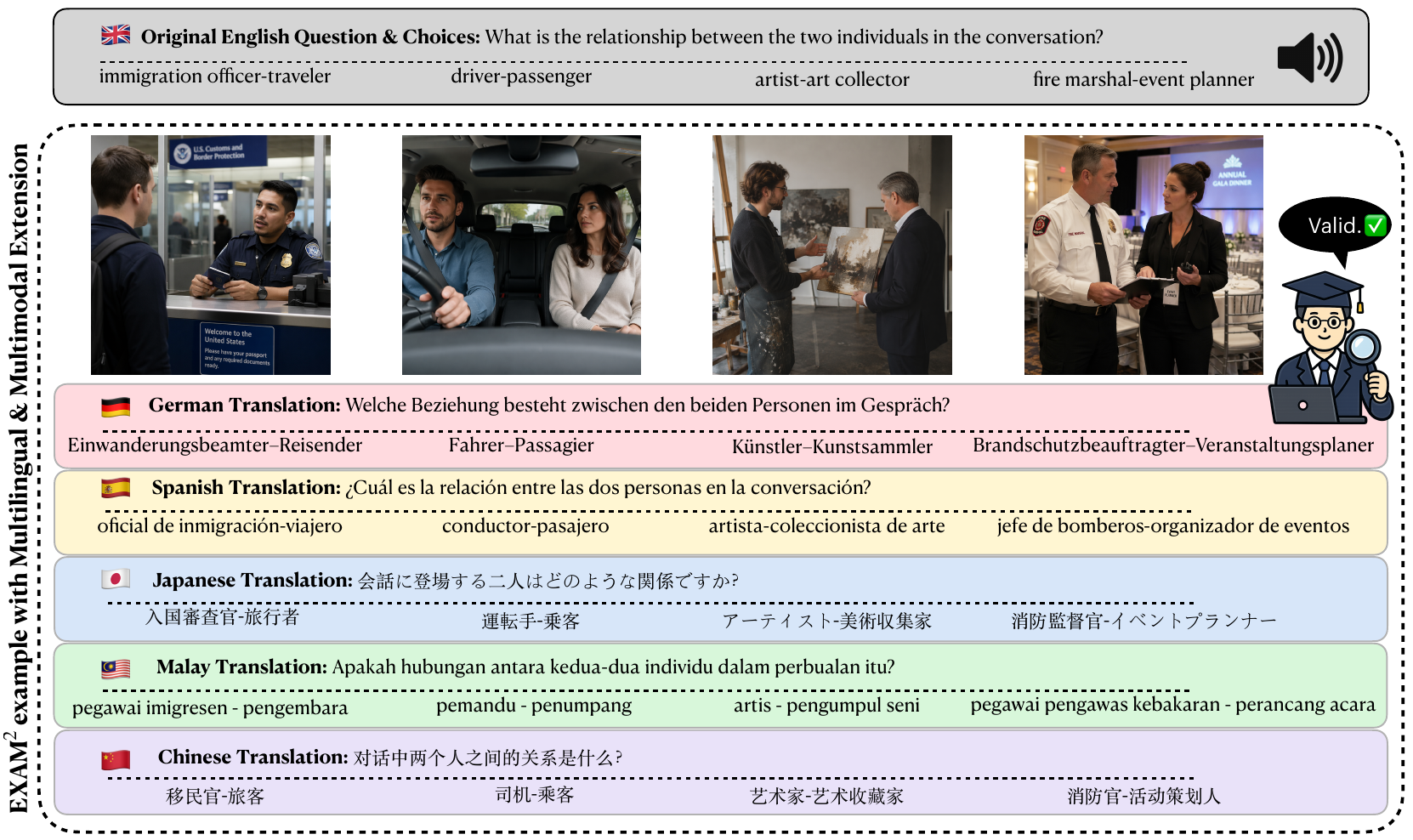}
  \caption{Overview of a multimodal and multilingual example from our EXAM$^2$ Benchmark.}
  \label{fig1}
\end{figure*}

\section{Introduction}
\label{sec:intro}
Recent advancements of \acp{LALM}~\cite{deshmukh2023pengi,chu2024qwen2,kong2024audio,ghosh2025audio,ghosh2026audio} has significantly expanded machine capabilities for understanding, reasoning over, and interacting with auditory signals~\cite{wang2025mmsu,pang2026erm}. 
To assess these capabilities, \ac{MCQ} tasks~\cite{he2025measuring} are being utilized across distinct audio modalities, including speech~\cite{zhao2024librisqa,huang2024dynamic}, sound~\cite{lipping2022clotho,ghosh2024compa}, and music~\cite{melechovsky-etal-2024-mustango,weck2024muchomusic}, to evaluate how models interpret and reason about audio content under various contexts and scenarios. 
Consequently, a wide array of benchmarks have emerged, including AudioBench~\cite{wang-etal-2025-audiobench}, MMAU~\cite{DBLP:conf/iclr/SakshiTKSSNDGM25}, and MMAR~\cite{DBLP:journals/corr/abs-2505-13032}. Yet, these benchmarks remain primarily audio-only and rarely investigate interactions between heterogeneous audio sources and visual grounding~\cite{yue2024mmmu,yue-etal-2025-mmmu,wang-etal-2025-omnieval,chen2025uno}. 

Beyond cross-modal limitations, restricted language coverage remains a critical bottleneck in current audio understanding evaluations. 
However, the semantic interpretation of audio content is frequently intertwined with linguistic and cultural contexts; for instance, identical lyrical slang can convey different meanings in diverse languages~\cite{werner-2023-english}. 
Yet, existing audio-understanding benchmarks largely rely on English-centric question-answer settings~\cite{drossos2020clotho, yang-etal-2024-air,DBLP:conf/iclr/SakshiTKSSNDGM25}, 
overlooking models' capabilities to navigate these shifting contexts. This discrepancy underscores the necessity for multilingual question-answer benchmarks to facilitate a broader assessment of audio comprehension across diverse language settings.



To address these gaps, we introduce EXAM$^2$, a benchmark for multilingual and multimodal audio understanding under a unified \ac{MCQ} evaluation framework (cf. \Cref{fig1}). 
EXAM$^2$ extends conventional audio-only benchmarks by jointly combining speech, sound, music, mixed-audio scenarios, and visual representations across six languages: English, German, Spanish, Japanese, Malay, and Chinese. 
Beyond measuring recognition accuracy, our benchmark probes multilingual reasoning, semantic grounding, and cross-modal comprehension in contemporary \acp{LALM}, \acp{MLLM}, and cascaded systems via \acp{LLM}. 
Finally, 
our proposed lightweight Gemma3n-EXAM$^2$ model achieves up to $+15.80\%$ average multilingual improvement ($+22.32\%$ on the speech category) and $16.50\%$ gains in multimodal settings compared to the baseline, demonstrating the model's effectiveness for advancing multilingual and multimodal audio \ac{MCQ} performance. 


Our contributions mainly include: 
(1) We propose the EXAM$^2$ benchmark, to our best knowledge, the first to address multilingual and multimodal audio understanding with $5,667$ \acp{MCQ}, $22,614$ image instances, and $135,684$ multilingual translated instances; 
(2) We introduce OmniLoRA, a straightforward omni-language uniform tuning based on LoRA, extensive experiments show the effectiveness of our method; 
(3) We present our multilingual Gemma3n-EXAM$^2$, which is a lightweight fusion approach with significant improvements across languages to foster future research in multilingual and multimodal domains; 
(4) We conduct extensive experiments using state-of-the-art \acp{LLM} and \acp{MLLM} on the EXAM$^2$ benchmark, offering a more holistic evaluation of audio intelligence and reasoning capabilities. 


\section{Related Work}
\label{sec:related}
\paragraph{Large Audio Language Models.}
Early works such as AudioCLIP~\cite{DBLP:conf/icassp/GuzhovRHD22} and CLAP~\cite{DBLP:conf/icassp/ElizaldeDIW23} focused on learning shared audio-text representations for retrieval and captioning tasks. 
Building on these foundations, \acp{LALM} such as Qwen2-Audio~\cite{chu2024qwen2}, Audio Flamingo~\cite{kong2024audio,ghosh2025audio,ghosh2026audio}, and SALMONN~\cite{tang2024salmonn} integrate audio encoders with \acp{LLM} to support instruction following and complex audio understanding. 
In parallel, general-purpose omni-models such as Qwen-omni~\cite{xu2025qwen3,team2026qwen3}, Ming-omni~\cite{ai2025ming} and Baichuan-omni~\cite{li2025baichuan} have shown strong cross-modal generalization despite not being specifically designed for audio tasks. 
While prior benchmarks mainly focus on evaluating \acp{LALM}, our work provides a comprehensive benchmark across \acp{LALM}, \acp{MLLM}, and cascaded \acp{LLM} that first transcribe audio before performing reasoning with (large) language or reasoning models. 

\paragraph{Multilingual and Multimodal Audio Understanding Benchmarks.} 
Previous benchmarks have advanced the evaluation of \acp{LALM} across speech, sound, and music understanding. 
Domain specific benchmarks such as LibriSQA~\cite{zhao2024librisqa} and Dynamic-SUPERB~\cite{huang2024dynamic} for speech-related tasks, 
Clotho-AQA~\cite{lipping2022clotho}, CompA~\cite{ghosh2024compa} for environmental sound reasoning~\cite{iyer-etal-2026-scenebench}, 
and MusicBench~\cite{melechovsky-etal-2024-mustango} and MuChoMusic~\cite{weck2024muchomusic} for music understanding. 
Broader benchmarks such as AudioBench~\cite{wang-etal-2025-audiobench}, AIR-Bench~\cite{yang-etal-2024-air}, MMAU~\cite{DBLP:conf/iclr/SakshiTKSSNDGM25}, MMAU-Pro~\cite{DBLP:conf/aaai/KumarSLLYARCPHE26}, and MMAR~\cite{DBLP:journals/corr/abs-2505-13032} further combine multiple audio domains and evaluate advanced auditory reasoning capabilities. 
However, existing benchmarks remain largely audio-only, with limited support for multimodal grounding or visual representations in audio understanding tasks. 
In addition, multilingual evaluation is still underexplored. 
Prior multilingual benchmarks such as BUFFET~\cite{asai-etal-2024-buffet} and MEGA~\cite{ahuja-etal-2023-mega} focus on text-based \acp{LLM} rather than the \acp{MLLM}, while the audio benchmark MMAR provides very limited multilingual audio and entirely lacks multilingual question-and-answer pairs.

\section{EXAM$^2$ Benchmark}
\label{sec:exam}

\begin{table}
  \centering
  \resizebox{.49\textwidth}{!}{
  \begin{tabular}{lccc|cccccc}
    \toprule
    \multirow{2}{*}{\textbf{Benchmark}} & \textbf{Text} & \textbf{Audio} & \textbf{Vision} & \multicolumn{6}{c}{\textbf{Language}}\\
    \cline{2-10}
    & Tr / Ca & Sp / So / Mu / Mix & Image & EN & DE & ES & JA & MS & ZH \\
    \midrule 
    Clotho &\ding{56} / \ding{52}&\ding{52} / \ding{52} / \ding{52} / \ding{56}&\ding{56}&\ding{52}&\ding{56}&\ding{56}&\ding{56}&\ding{56}&\ding{56} \\
    Clotho-AQA &\ding{56} / \ding{56}&\ding{56} / \ding{52} / \ding{56} / \ding{56}&\ding{56}&\ding{52}&\ding{56}&\ding{56}&\ding{56}&\ding{56}&\ding{56} \\
    LibriSQA &\ding{56} / \ding{56}&\ding{52} / \ding{56} / \ding{56} / \ding{56}&\ding{56}&\ding{52}&\ding{56}&\ding{56}&\ding{56}&\ding{56}&\ding{56} \\
    Dynamic-SUPERB & \ding{56} / \ding{56}&\ding{52} / \ding{56} / \ding{56} / \ding{56}&\ding{56}&\ding{52}&\ding{56}&\ding{56}&\ding{56}&\ding{56}&\ding{56} \\
    MusicBench &\ding{56} / \ding{56}&\ding{56} / \ding{56} / \ding{52} / \ding{56}&\ding{56}&\ding{52}&\ding{56}&\ding{56}&\ding{56}&\ding{56}&\ding{56} \\
    MuChoMusic &\ding{56} / \ding{56}&\ding{56} / \ding{56} / \ding{52} / \ding{56}&\ding{56}&\ding{52}&\ding{56}&\ding{56}&\ding{56}&\ding{56}&\ding{56} \\
    AIR-Bench &\ding{56} / \ding{56}&\ding{52} / \ding{52} / \ding{52} / \ding{52}&\ding{56}&\ding{52}&\ding{56}&\ding{56}&\ding{56}&\ding{56}&\ding{56} \\
    AudioSetCaps &\ding{56} / \ding{52}&\ding{52} / \ding{52} / \ding{52} / \ding{56}&\ding{56}&\ding{52}&\ding{56}&\ding{56}&\ding{56}&\ding{56}&\ding{56} \\
    AudioMCQ &\ding{56} / \ding{56}&\ding{52} / \ding{52} / \ding{52} / \ding{56}&\ding{56}&\ding{52}&\ding{56}&\ding{56}&\ding{56}&\ding{56}&\ding{56} \\
    MMAR &\ding{56} / \ding{56}&\ding{52} / \ding{52} / \ding{52} / \ding{52}&\ding{56}&\ding{52}&\ding{52}&\ding{56}&\ding{52}&\ding{56}&\ding{52} \\
    MMAU &\ding{56} / \ding{56}&\ding{52} / \ding{52} / \ding{52} / \ding{56}&\ding{56}&\ding{52}&\ding{56}&\ding{56}&\ding{56}&\ding{56}&\ding{56} \\
    MMAU-Pro &\ding{56} / \ding{56}&\ding{52} / \ding{52} / \ding{52} / \ding{56}&\ding{56}&\ding{52}&\ding{56}&\ding{56}&\ding{56}&\ding{56}&\ding{56} \\
    MMMU &\ding{56} / \ding{56}&\ding{56} / \ding{56} / \ding{56} / \ding{56}&\ding{52}&\ding{52}&\ding{56}&\ding{56}&\ding{56}&\ding{56}&\ding{56} \\
    MMAU-Pro &\ding{56} / \ding{56}&\ding{56} / \ding{56} / \ding{56} / \ding{56}&\ding{52}&\ding{52}&\ding{56}&\ding{56}&\ding{56}&\ding{56}&\ding{56} \\
    \midrule 
    \textbf{EXAM$^2$ (Ours)}&\ding{52} / \ding{52}&\ding{52} / \ding{52} / \ding{52} / \ding{52}&\ding{52}&\ding{52}&\ding{52}&\ding{52}&\ding{52}&\ding{52}&\ding{52}\\
    \bottomrule
  \end{tabular}
  }
  \caption{\label{difference}
    The list of existing multimodal and multilingual datasets show the novelty of our EXAM$^2$ benchmark. 
    Tr, Ca denotes Transcript and Caption, respectively. Sp / So / Mu / Mix stands for Speech, Sound, Music, and their Mix.
  }
\end{table}

\subsection{Overview}
Our EXAM$^2$ introduces multilingual and multimodal audio understanding with visual representations, supporting six languages (DE, ES, JA, MS, ZH, and EN) to enable richer cross-modal and cross-lingual reasoning (cf. \Cref{difference}). 
Dataset statistics for our proposed EXAM$^2$ benchmark are given in \Cref{statistic}. 
The benchmark consists of two main subsets: EXAM$^2$-train and EXAM$^2$-test, with a total of $5,667$ audio instances, $135,684$ multilingual language instances, and $22,614$ visual answer choices. 
EXAM$^2$-train, recreated and filtered from MMAR and Clotho, includes $4,669$ questions covering four audio domains (speech, sound, music, and mix) with $18,676$ visual choices. 
EXAM$^2$-test, restructured and annotated from MMAU-test-mini, contains $998$ questions across three audio domains (speech, sound, and music) with a balanced category distribution and a total of $3,938$ visual choices. 

\begin{table}
  \centering
  \resizebox{.48\textwidth}{!}{
  \begin{tabular}{llr}
    \toprule 
    \textbf{Benchmark}&\textbf{Statistics}&\textbf{Number}\\
    \midrule 
    \multirow{7}{*}{EXAM$^2$-test} & Question & 998\\
    & Audio Domains & 3 \\
    & Domain Categories & Speech / Sound / Music \\
    & Category Distribution & 333:333:332 (1:1:1) \\
    & Visual Choices & 3,938 \\
    & Languages & EN / DE / ES / JA / MS / ZH \\
    & Multilingual Instances & 23,628 \\
    \midrule
    \multirow{7}{*}{EXAM$^2$-train} & Question & 4669\\
    & Audio Domains & 4 \\
    & Domain Categories & Speech / Sound / Music / Mix\\
    & Category Distribution & 396:3737:243:293 \\
    & Visual Choices & 18,676 \\
    & Languages & EN / DE / ES / JA / MS / ZH \\
    & Multilingual Instances & 112,056 \\
    \midrule 
    \multirow{10}{*}{EXAM$^2$} & Question & 5667\\
    & Audio Domains & 4 \\
    & Domain Categories & Speech / Sound / Music / Mix\\
    & Category Distribution & 729:4070:575:293 \\
    & Visual Choices & 22,614 \\
    & Languages & EN / DE / ES / JA / MS / ZH \\
    & Multilingual Instances & 135,684\\
    \cline{2-3}
    & Average Question Length & 10.7 words\\
    & Average Option Length & 3.2 words\\
    & Average Audio Duration & 12.9 sec\\
    \bottomrule
  \end{tabular}
  }
  \caption{\label{statistic}
    Dataset statistics of our EXAM$^2$ benchmark.
  }
\end{table}

\subsection{Dataset Construction}

\begin{figure*}[ht]
  \includegraphics[width=\linewidth]{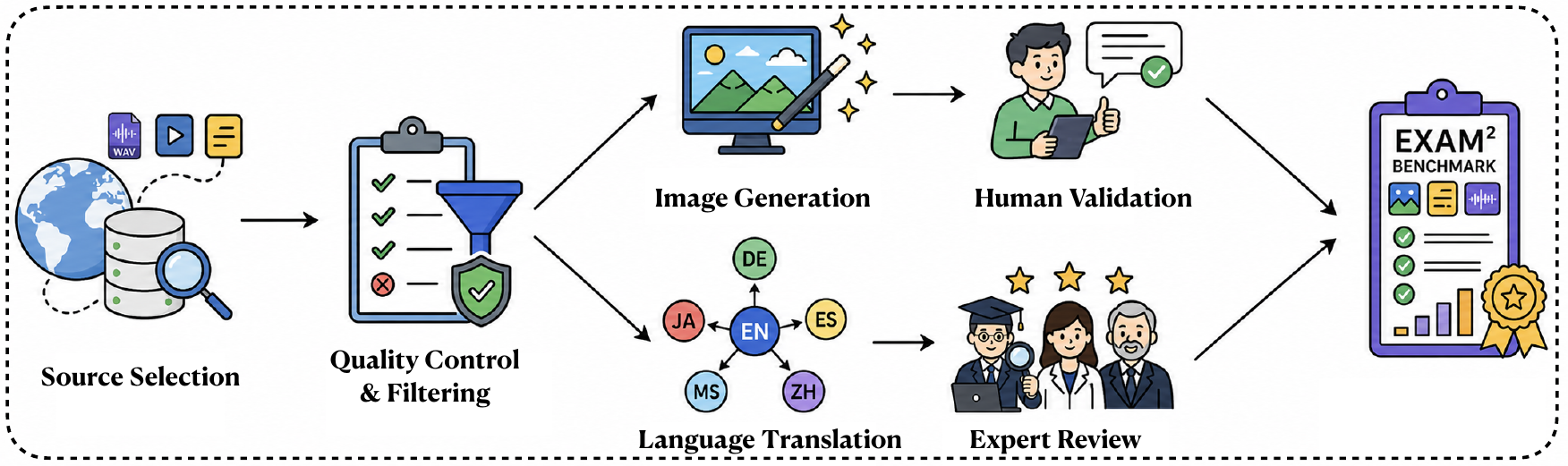}
  \caption{EXAM$^2$ Benchmark Construction Pipeline.}
  \label{fig2}
\end{figure*}

\Cref{fig2} illustrates the construction pipeline of our EXAM$^2$ benchmark with seven key steps.  

\paragraph{Source Selection.}
We begin by collecting diverse audio corpora spanning speech, music, and environmental sounds, prioritizing real-world recordings over synthetic data to ensure ecological validity. 
Specifically, we curate four representative datasets: MMAU-test-mini~\cite{DBLP:conf/iclr/SakshiTKSSNDGM25}, MMAR~\cite{DBLP:journals/corr/abs-2505-13032}, Clotho~\cite{drossos2020clotho}, and AudioMCQ-Clotho~\cite{he2025measuring} under \texttt{Creative Commons} license to ensure a strong foundation for task development. 

\paragraph{Quality Control and Filtering.}
To ensure benchmark quality, all collected audio samples, textual questions, choices, and answers undergo manual inspection by the authors. 
For MMAU-test-mini, we remove two music-category samples that fail to meet the quality standards and eliminate duplicate answer choices within the same question, resulting in a final test set of $998$ instances with $3,938$ answer choices. 
For MMAR, we retrieve audio samples from the provided URLs and discard $129$ out of $1,000$ instances due to inaccessible or broken links, yielding $871$ valid audio instances with $3,484$ answer choices. 
For the Clotho benchmark, we derive multiple-choice questions from AudioMCQ-Clotho, obtaining $3,798$ audios with $15,192$ choices. 

\paragraph{Image Generation.}
To enable multimodal evaluation, we generate visual counterparts for all answer choices in EXAM$^2$. For the smaller subsets, EXAM$^2$-MMAU and EXAM$^2$-MMAR, we employ Stable Diffusion~\cite{rombach2022high} for image generation under \texttt{CreativeML Open RAIL-M} license. 
For EXAM$^2$-Clotho, visual answer choices are generated using GPT-image-2.\footnote{\url{https://openai.com/index/introducing-chatgpt-images-2-0/}} 

\paragraph{Human Validation.}
All textual questions and generated visual candidates are manually reviewed by the experts to ensure semantic relevance, clarity, and overall quality. 
When generated images are judged to be misleading, ambiguous, or insufficiently aligned with the corresponding answer choice, we manually modified 25.05\% of the generated images to real-world screenshots to ensure the visual quality. 
Representative examples across categories are provided in \Cref{sec:appendix:data:visual}. 

\paragraph{Language Translation.}
To support multilingual evaluation, we translate the original English questions and answer choices into five target languages: German (DE), Spanish (ES), Japanese (JA), Malay (MS), and Chinese (ZH). 
We employ GPT-5-mini\footnote{\url{https://openai.com/index/introducing-gpt-5/}} for translation, using carefully designed prompts (See \Cref{sec:appendix:prompt}) to preserve semantic fidelity and task consistency across languages. 
Domain-specific terminology related to audio understanding is further reviewed and manually refined where necessary to ensure accurate translation and conceptual equivalence. 

\paragraph{Expert Review.}
Translated questions and answer choices are reviewed by either native speakers of the target languages or experienced language users with more than five years of advanced proficiency. 
The review process focuses on semantic correctness, linguistic fluency, and preservation of task intent across languages, ensuring high-quality multilingual question-answer pairs across all benchmark languages. 
The manual edit rates of translation are 3.01\% (DE), 2.30\% (ES), 2.81\% (JA), 1.10\% (MS), and 5.31\% (ZH). An inter-annotator agreement of Cohen’s kappa achieves $0.9830$, indicating almost perfect agreement. 

\paragraph{EXAM$^2$ Finalization.}
The final EXAM$^2$ benchmark comprises $5,667$ audio instances, $135,684$ multilingual language instances, and $22,614$ visual answer choices. 
The benchmark is designed to provide balanced coverage of speech, sound, and music understanding across six languages (EN, DE, ES, JA, MS, and ZH), enabling comprehensive evaluation of multimodal and multilingual audio understanding.

\section{Methodology}
\label{sec:method}

\subsection{Task Formulation}
\label{sec:method:task}
We address multimodal \ac{MCQ} over audio and visual inputs.
Each instance is a tuple $(\mathbf{a},\,\mathcal{V},\,q,\,\mathcal{C},\,y^*)$, where $\mathbf{a} \in \mathbb{R}^{T}$ is a raw audio waveform of $T$ samples, $\mathcal{V} = \{v_1, \dots, v_K\}$ is a set of $K$ candidate images (one per choice), $q$ is a natural-language question, $\mathcal{C} = \{c_1, \dots, c_K\}$ is the set of candidate answers, and $y^* \in \mathcal{C}$ is the ground-truth answer.
The model must identify $y^*$ by jointly reasoning over both modalities. 

\subsection{Multimodal Input Representation}
\label{sec:method:input}
We build on Gemma3n-E4B\footnote{\url{https://huggingface.co/google/gemma-3n-E4B-it}}, a \ac{MLLM} that encodes audio, images, and text within a unified transformer.

\paragraph{Audio.}
The waveform $\mathbf{a}$ is resampled to 16\,kHz and converted to a spectrogram $\mathbf{S} \in \mathbb{R}^{F \times L}$ ($F$ frequency bins, $L$ frames). 
A convolutional audio encoder projects $\mathbf{S}$ into a sequence of $d$-dimensional hidden states $\mathbf{H}^{(a)} \in \mathbb{R}^{L' \times d}$, which are prepended to the token sequence as soft audio tokens.

\paragraph{Images.}
For each choice $c_k$, a corresponding image $v_k \in \mathbb{R}^{H\times W \times 3}$ is encoded into patch embeddings $\mathbf{H}^{(v_k)} \in \mathbb{R}^{N_p \times d}$, where $N_p$ is the number of patches. 
The image embeddings for all $K$ choices are concatenated and interleaved with textual tokens. 

\paragraph{Multimodal Fusion.}
The combined input sequence fed to the language-model backbone is
\begin{equation}
  \mathbf{X} =
    \bigl[\,\mathbf{H}^{(v_1)},\ldots,\mathbf{H}^{(v_K)},\;
           \mathbf{H}^{(a)},\;
           \mathbf{e}_q,\;
           \mathbf{e}_{\mathcal{C}}\,\bigr],
  \label{eq:fusion}
\end{equation}
where $\mathbf{e}_q$ and $\mathbf{e}_{\mathcal{C}}$ are the token embeddings of the question and choice text, respectively.
No modality-specific fusion layer is added; cross-modal alignment is learned entirely through self-attention. 

\subsection{Parameter-Efficient Fine-Tuning via LoRA}
\label{sec:method:lora}
Full fine-tuning of all model parameters $\theta$ is computationally prohibitive.
Following \citet{hu2022lora}, we freeze $\theta$ and inject trainable low-rank update matrices into the attention and feed-forward layers. 
We fine-tune the model with a cross-entropy loss restricted to the answer token.
Given the multimodal input $\mathbf{X}$ (Eq.~\ref{eq:fusion}), the model is trained to predict the single-digit index $y^* \in \{1,\ldots,K\}$ identifying the correct choice:
\begin{equation}
  \mathcal{L}(\theta_{\mathrm{LoRA}})
    = -\log P_\theta\!\left(y^* \mid \mathbf{X}\right).
  \label{eq:loss}
\end{equation}

After training, the LoRA adapters are merged back into the base weights and saved as a single self-contained model for inference. 
To support multilingual evaluation, we propose omni-language uniform tuning (OmniLoRA), which applies LoRA to the backbone model by uniformly sampling one of six languages $\ell \in \{\texttt{EN, DE, ES, JA, MS, ZH}\}$ for each training instance per epoch: 
\begin{equation}
  \mathcal{L}_{\mathrm{Omni}}(\theta_{\mathrm{LoRA}})
    = -\,\mathbb{E}_{\ell \sim \mathcal{L}}\!\left[
        \log P_\theta\!\left(y^* \mid \mathbf{X}^{(\ell)}\right)
      \right],
  \label{eq:omnilora}
\end{equation}
where $\mathbf{X}^{(\ell)}$ denotes the fused multimodal sequence (Eq.~\ref{eq:fusion}) with text rendered in language $\ell$.
This design keeps pre-trained multimodal representations intact while efficiently adapting the model to the target task across six languages.

\section{Evaluation}
\label{sec:evaluation}

\subsection{Experimental Setup} 
\label{sec:evaluation:setup} 
We evaluate a diverse set of closed- and open-source models on the proposed EXAM$^2$ benchmark. 
Our results can be easily reproduced using the provided codebase and evaluation scripts. 
Closed-source models and Phi-4-multimodal-instruct are conducted using the Azure OpenAI API, while the remaining open-source models are evaluated using Hugging Face library. 
The supervised fine-tuning on our EXAM$^2$-train allocates four NVIDIA A40 GPUs. 
For all models and languages, we use the same set of prompts, detailed in Appendix~\ref{sec:appendix:prompt}, across audio domains (speech, sound, and music) to ensure a fair comparison. 
Note that GPT-4o-transcribe\footnote{\url{https://developers.openai.com/api/docs/models/gpt-4o-transcribe}} is deployed for audio transcription tasks.

\subsection{Baseline} 
\label{sec:evaluation:model} 
Closed- and open-source models are evaluated on the EXAM$^2$-test, in zero-shot settings. 
For closed-source models, we evaluate GPT-4o-audio\footnote{\url{https://developers.openai.com/api/docs/models/gpt-4o-audio-preview}} and GPT-4o-mini-audio\footnote{\url{https://developers.openai.com/api/docs/models/gpt-4o-mini-audio-preview}}. 
GPT-4o\footnote{\url{https://developers.openai.com/api/docs/models/gpt-4o}} is leveraged as a \ac{MLLM} with image input, while GPT-5-mini\footnote{\url{https://openai.com/gpt-5/}} and GPT-5.2\footnote{\url{https://openai.com/index/introducing-gpt-5-2/}} are evaluated as text-only models to assess the impact of multimodal grounding and multilingual competence. 
Phi-4-multimodal-instruct variant\footnote{\url{https://huggingface.co/microsoft/Phi-4-multimodal-instruct}}, along with other models such as Qwen-2-audio-7B-instruct\footnote{\url{https://huggingface.co/Qwen/Qwen2-Audio-7B-Instruct}} Qwen-2.5-omni-4-bit-GPTQ\footnote{\url{https://huggingface.co/Qwen/Qwen2.5-Omni-7B}} and Gemma-3n-E4B\footnote{\url{https://huggingface.co/google/gemma-3n-E4B-it}}, are evaluated as open-source alternatives. 

\subsection{Results and Analysis} 
\label{sec:evaluation:result}

\begin{table}
  \small
  \centering
  \setlength{\tabcolsep}{2.5pt}
  \renewcommand{\arraystretch}{0.92}
  \begin{adjustbox}{max width=\linewidth, max totalheight=0.88\textheight, keepaspectratio}
  \begin{tabular}{@{}lcc|cccc@{}}
    \toprule
    \multirow{2}{*}{\textbf{Model}} 
    & \multicolumn{2}{c|}{\textbf{Input Modality}} 
    & \multicolumn{4}{c}{\textbf{Average (\%)}} \\
    & Image & Audio 
    & Speech & Sound & Music & Avg. \\
    \midrule
    \textbf{\textit{Closed-source}} & & & & & & \\ 
    \midrule
    GPT-4o-mini-audio 
    & \ding{56} & \ding{52} 
    & 73.67 & 61.01 & 57.13 & 63.94 \\

    GPT-4o-audio 
    & \ding{56} & \ding{52} 
    & 77.93 & 73.17 & 70.38 & 73.83 \\

    GPT-4o$^\dagger$ 
    & \ding{52} & \ding{51} 
    & 66.67 & 69.17 & 57.13 & 64.32 \\

    GPT-5-mini$^\dagger$ 
    & \ding{56} & \ding{51} 
    & 71.82 & 51.25 & 54.29 & 59.12 \\

    GPT-5-2$^\dagger$ 
    & \ding{56} & \ding{51} 
    & 71.47 & 50.25 & 50.10 & 57.27 \\

    \midrule
    \textbf{\textit{Open-source}} & & & & & & \\ 
    \midrule
    Phi-4-multimodal-instruct 
    & \ding{56} & \ding{52} 
    & \underline{59.91} & 54.47 & 55.02 & 56.47 \\

    Phi-4-multimodal-instruct$^\dagger$ 
    & \ding{56} & \ding{51} 
    & 47.05 & 60.81 & 56.21 & 54.69 \\

    Phi-4-multimodal-instruct$^\dagger$ 
    & \ding{52} & \ding{51} 
    & 51.30 & 59.91 & 44.08 & 51.76 \\

    Qwen-2-audio&\ding{56}&\ding{52}
    & 34.29 & 34.33 & 40.36 & 36.33 \\

    Qwen-2.5-omni$^\dagger$ 
    & \ding{56} & \ding{51} 
    & 53.60 & 64.17 & \underline{56.52} & 58.10 \\

    Qwen-2.5-omni$^\dagger$ 
    & \ding{52} & \ding{51} 
    & 53.25 & 61.11 & 55.37 & 56.58 \\

    Qwen-2.5-omni 
    & \ding{56} & \ding{52} 
    & 39.64& 49.90 & 32.58 & 40.71 \\

    Qwen-2.5-omni 
    & \ding{52} & \ding{52} 
    & 52.20 & 61.61 & 54.97 & 56.26 \\

    Gemma3n$^\dagger$ 
    & \ding{56} & \ding{51} 
    & 58.41 & \textbf{67.12} & 54.22 & \underline{59.91} \\

    Gemma3n$^\dagger$ 
    & \ding{52} & \ding{51} 
    & 57.16 & \underline{66.37} & 53.46 & 59.00 \\

    Gemma3n 
    & \ding{52} & \ding{56} 
    & 42.14 & 48.70 & 46.64 & 45.83 \\

    Gemma3n 
    & \ding{56} & \ding{52} 
    & 42.84 & 43.04 & 48.34 & 44.74 \\

    Gemma3n 
    & \ding{52} & \ding{52} 
    & 41.99 & 47.35 & 46.99 & 45.44 \\

    \textbf{Gemma3n-EXAM$^2$ (Ours)} 
    & \ding{52} & \ding{52} 
    & \textbf{64.31} & 59.26 & \textbf{60.14} & \textbf{61.24} \\

    \midrule
    Random guess & \ding{52} & \ding{52} & 27.18 & 24.97 & 25.02 & 25.72 \\ 
    Human ceiling (Ours) & \ding{52} & \ding{52} & 91.59 & 84.08 & 65.66 & 80.44 \\
    \bottomrule
  \end{tabular}
  \end{adjustbox}
  \caption{\label{mmau-exam-average}
    Average multilingual and multimodal evaluation results across the six languages over Speech, Sound, and Music. 
    \textbf{Bold} and \underline{underline} indicate the best and second best performance for each category among open-source models only.
    $^\dagger$ shows cascaded input of transcript and/or caption. 
    \ding{52} and \ding{51} are marked to differentiate the original audio waveform and textual input, respectively. \ding{56} marks the unexploited modality. 
  }
\end{table}

Table~\ref{mmau-exam-average} summarizes the average performance on the EXAM$^2$ benchmark across three audio domains (speech, sound, and music) and six languages (DE, EN, ES, JA, MS, and ZH). Among closed-source models, GPT-4o-audio achieves the strongest overall performance with an average score of 73.83\%, consistently leading in Speech, Sound, and Music. Focusing on open-source models, our proposed Gemma3n-EXAM$^2$ obtains the best overall average of 61.24\% (with a significant improvement over Gemma3n, by paired t-test, $p < 0.05$), and achieves the highest scores in both Speech and Music with 64.31\% and 60.14\%, respectively. 
However, Gemma3n$^\dagger$ (given the cascaded textual input) remains stronger in Sound, achieving 67.12\% compared with 59.26\% for Gemma3n-EXAM$^2$, indicating that our model is particularly effective for speech- and music-related understanding while leaving room for further improvement on sound-centric reasoning.

Meanwhile, \Cref{mmau-exam} presents our results across all six languages. 
Overall, our proposed Gemma3n-EXAM$^2$ model, compared to baseline with the same modalities input, demonstrates competitive performance across all languages and audio categories, with a notable increase of $26.43\%$ in the speech domain for German and $+22.32\%$ accuracy gain for all languages on average. 
A global improvement of $11.91\%$ and $13.15\%$ in the sound and music domain respectively is observed, as both sound and music benefit from the multimodal grounding provided by the visual answer choices. 
By utilizing our OmniLoRA approach on Gemma3n-E4B with our multilingual and multimodal training data, we can significantly boost the performance of the model across all languages and audio domains, demonstrating the high quality of our benchmark for improving multilingual and multimodal audio understanding.

\begin{table*}[ht]
  \small
  \centering
  \resizebox{\textwidth}{!}{
  \begin{tabular}{lcc|cc|cc|cc}
    \toprule
    \multirow{2}{*}{\textbf{Model}}& \multicolumn{2}{c}{\textbf{Input Modality}}&\multicolumn{2}{c}{\textbf{German}}&\multicolumn{2}{c}{\textbf{English}}&\multicolumn{2}{c}{\textbf{Spanish}}\\
    & Image & Audio
    & Speech / Sound / Music & Avg.
    & Speech / Sound / Music & Avg.
    & Speech / Sound / Music & Avg.\\
    \midrule
    \textbf{\textit{Closed-source}} &\\ \midrule 
    GPT-4o-mini-audio&\ding{56}&\ding{52}&75.08 / 59.46 / 56.63   &63.72&74.77 / 59.16 / 57.53&63.82&71.47 / 62.46 / 59.34&64.42\\
    GPT-4o-audio&\ding{56}&\ding{52}&78.98 / 70.87 / 70.18        &73.34&77.18 / 72.67 / 71.39&73.75&75.98 / 75.08 / 72.29&74.45\\
    GPT-4o$^\dagger$&\ding{52}&\ding{51}&65.77 / 69.07 / 57.23    &64.02&68.77 / 70.57 / 59.64&66.33&67.57 / 68.17 / 59.04&64.93\\
    GPT-5-mini$^\dagger$&\ding{56}&\ding{51}&72.37 / 51.05 / 54.22&59.21&72.67 / 55.56 / 55.72&61.32&72.67 / 50.45 / 55.12&59.41\\
    GPT-5-2$^\dagger$&\ding{56}&\ding{51}&71.17 / 48.05 / 51.51   &56.91&72.67 / 51.95 / 50.00&58.21&69.07 / 54.05 / 51.51&58.21\\
    \midrule 
    \textbf{\textit{Open-source}} &\\ \midrule 
    Phi-4-multimodal-instruct&\ding{56}&\ding{52}&\underline{61.56} / 57.66 / 53.31       &57.51&\textbf{65.17} / 65.57 / \textbf{64.16}&\textbf{64.97}&\underline{61.86} / 55.56 / 56.63&58.02\\
    Phi-4-multimodal-instruct$^\dagger$&\ding{56}&\ding{51}&45.95 / 61.86 / \textbf{59.94}&55.92&50.75 / 65.47 / \underline{63.25}      &59.82&49.55 / 64.56 / \underline{56.93}         &57.01\\
    Phi-4-multimodal-instruct$^\dagger$&\ding{52}&\ding{51}&51.65 / 63.66 / 44.28         &53.20&54.95 / 62.76 / 50.90                  &56.20&50.75 / 59.76 / 44.88                     &51.80\\
    Qwen-2-audio&\ding{56}&\ding{52}&28.23 / 34.83 / 39.46                                &34.17&38.74 / 41.74 / 47.89                  &42.79&35.14 / 32.43 / 43.37                     &36.98\\
    Qwen-2.5-omni$^\dagger$&\ding{56}&\ding{51}&51.65 / 65.17 / 58.43                     &58.42&58.86 / \underline{68.17} / 58.73      &61.92&53.45 / 63.36 / 56.33                     &57.71\\
    Qwen-2.5-omni$^\dagger$&\ding{52}&\ding{51}&54.65 / 62.46 / \underline{58.73}         &58.61&58.56 / 67.87 / 59.04                  &61.82&53.15 / 60.06 / 55.42                     &56.21\\
    Qwen-2.5-omni&\ding{56}&\ding{52}&41.44 / 51.35 / 34.94                               &42.58&39.34 / 48.95 / 30.42                  &39.57&40.54 / 48.35 / 31.33                     &40.07\\
    Qwen-2.5-omni&\ding{52}&\ding{52}&51.05 / 61.26 / 57.83                               &56.71&55.86 / 66.97 / 58.73                  &60.52&51.65 / 61.86 / 55.42                     &56.31\\
    Gemma3n$^\dagger$&\ding{56}&\ding{51}&56.76 / \textbf{69.67} / 52.41                  &\underline{59.61}&60.06 / 66.67 / 56.93      &61.22&58.86 / \textbf{68.47} / 53.01            &\underline{60.11}\\
    Gemma3n$^\dagger$&\ding{52}&\ding{51}&55.86 / \underline{69.07} / 53.61               &59.51&58.86 / \textbf{70.57} / 55.42         &61.62&56.16 / \underline{65.17} / 55.12         &58.82\\
    Gemma3n&\ding{52}&\ding{56}&41.44 / 48.35 / 46.39                                     &45.39&42.94 / 51.65 / 48.19                  &47.59&39.94 / 47.75 / 45.48                     &44.39\\
    Gemma3n&\ding{56}&\ding{52}&41.74 / 45.05 / 47.29                                     &44.69&47.75 / 43.84 / 51.81                  &47.80&47.45 / 41.44 / 48.19                     &45.69\\
    Gemma3n &\ding{52}&\ding{52}&39.94 / 48.05 / 46.39                                    &44.79&43.24 / 48.35 / 48.49                  &46.69&40.24 / 48.95 / 47.89                     &45.69\\
    \midrule
    \textbf{Gemma3n-EXAM$^2$ (Ours)}&\ding{52}&\ding{52}&\textbf{66.37} / 62.76 / \textbf{59.94}&\textbf{63.02}&\underline{64.56} / 62.16 / 59.94&\underline{62.22}&\textbf{64.26} / 58.86 / \textbf{61.14}&\textbf{61.42}\\
    \bottomrule
    \toprule
    \multirow{2}{*}{\textbf{Model}}& \multicolumn{2}{c}{\textbf{Input Modality}}&\multicolumn{2}{c}{\textbf{Japanese}}&\multicolumn{2}{c}{\textbf{Malay}}&\multicolumn{2}{c}{\textbf{Chinese}}\\
    & Im & Au 
    & Speech / Sound / Music & Avg. 
    & Speech / Sound / Music & Avg. 
    & Speech / Sound / Music & Avg.\\
    \midrule
    \textbf{\textit{Closed-source}} &\\ \midrule 
    GPT-4o-mini-audio&\ding{56}&\ding{52}&74.47 / 62.16 / 56.02   &64.22&71.17 / 60.06 / 56.33&62.52&75.08 / 62.76 / 56.93&64.92\\
    GPT-4o-audio&\ding{56}&\ding{52}&79.28 / 74.17 / 68.07        &73.84&78.68 / 73.27 / 69.88&73.94&77.48 / 72.97 / 70.48&73.64\\
    GPT-4o$^\dagger$&\ding{52}&\ding{51}&66.97 / 68.47 / 55.42    &63.62&63.36 / 69.07 / 55.72&62.72&67.57 / 69.67 / 55.72&64.32\\
    GPT-5-mini$^\dagger$&\ding{56}&\ding{51}&71.77 / 48.65 / 53.92&58.11&69.97 / 51.65 / 54.82&58.81&71.47 / 50.15 / 51.96&57.86\\
    GPT-5-2$^\dagger$&\ding{56}&\ding{51}&72.37 / 48.05 / 48.19   &56.20&70.57 / 52.25 / 50.90&57.91&72.97 / 47.15 / 48.49&56.20\\
    \midrule 
    \textbf{\textit{Open-source}} &\\ \midrule 
    Phi-4-multimodal-instruct&\ding{56}&\ding{52}&59.16 / 51.05 / 51.81                      &54.01&51.05 / 44.14 / 49.70                                          &48.30&\underline{60.66} / 52.85 / 54.52            &56.01\\
    Phi-4-multimodal-instruct$^\dagger$&\ding{56}&\ding{51}&48.95 / 59.46 / \underline{54.22}&54.21&42.04 / 54.05 / 49.01                                          &48.37&45.05 / 59.46 / 53.92                        &52.81\\
    Phi-4-multimodal-instruct$^\dagger$&\ding{52}&\ding{51}&50.15 / 58.56 / 42.17            &50.29&46.25 / 57.06 / 39.16                                          &47.49&54.05 / 57.66 / 43.07                        &51.59\\
    Qwen-2-audio&\ding{56}&\ding{52}&29.73 / 26.73 / 34.34                                   &30.27&36.04 / 29.73 / 31.93                                          &32.57&37.84 / 40.54 / 45.18                        &41.19\\
    Qwen-2.5-omni$^\dagger$&\ding{56}&\ding{51}&51.65 / 63.36 / 53.01                        &56.01&48.65 / 59.16 / 54.82                                          &54.21&57.36 / \textbf{65.77} / \textbf{57.83}      &\textbf{60.32}\\
    Qwen-2.5-omni$^\dagger$&\ding{52}&\ding{51}&48.95 / 58.56 / 50.00                        &52.50&48.05 / 56.46 / 53.92                                          &52.81&56.16 / 61.26 / 55.12                        &57.51\\
    Qwen-2.5-omni&\ding{56}&\ding{52}&38.74 / 48.35 / 33.43                                  &40.17&38.14 / 52.55 / 32.53                                          &41.07&39.64 / 49.85 / 32.83                        &40.77\\
    Qwen-2.5-omni&\ding{52}&\ding{52}&50.15 / 59.16 / 48.19                                  &52.50&46.55 / 56.46 / 52.71                                          &51.91&57.96 / 63.96 / 56.93                        &\underline{59.62}\\
    Gemma3n$^\dagger$&\ding{56}&\ding{51}&59.76 / \textbf{69.07} / 53.92                     &\textbf{60.92}&\underline{54.65} / \textbf{64.56} / \underline{56.33}&\underline{58.51}&60.36 / \underline{64.26} / 52.71&59.11\\
    Gemma3n$^\dagger$&\ding{52}&\ding{51}&\underline{60.06} / \underline{65.77} / 49.70      &58.51&53.45 / \underline{63.66} / 55.12                              &57.41&58.56 / 63.96 / 51.81                        &58.11\\
    Gemma3n&\ding{52}&\ding{56}&43.24 / 46.85 / 47.29                                        &45.79&41.44 / 47.45 / 47.59                                          &45.49&43.84 / 50.15 / 44.88                        &46.29\\
    Gemma3n&\ding{56}&\ding{52}&41.74 / 42.34 / 45.78                                        &43.29&34.53 / 43.24 / 49.10                                          &42.29&43.84 / 42.34 / 47.89                        &44.69\\
    Gemma3n &\ding{52}&\ding{52}&43.24 / 45.05 / 46.08                                       &44.79&42.04 / 45.95 / 47.89                                          &45.29&43.24 / 47.75 / 45.18                        &45.39\\
    \midrule
    \textbf{Gemma3n-EXAM$^2$ (Ours)}&\ding{52}&\ding{52}&\textbf{63.36} / 56.76 / \textbf{59.64}&\underline{59.92}&\textbf{63.06} / 57.96 / \textbf{62.65}&\textbf{61.22}&\textbf{64.26} / 57.06 / \underline{57.53}&\underline{59.62}\\
    \bottomrule
  \end{tabular}
  }
  \caption{\label{mmau-exam}
    Multilingual and multimodal evaluation results (accuracy in \%) on the EXAM$^2$ test set. 
    \textbf{Bold} represents the best performance, while \underline{underline} indicates the second best performance for each category and each language within open-source models.
    $^\dagger$ shows cascaded input of transcript and/or caption. We also mark \ding{52} and \ding{51} to differentiate the original audio waveform and textual input, respectively. \ding{56} marks the unexploited modality. 
  }
\end{table*}

\paragraph{Image modality helps audio understanding.}
For all languages, we notice a significant improvement by incorporating visual context into audio understanding, particularly for native multimodal settings. 
Models with joint image–audio capability frequently outperform their audio-only or cascaded counterparts on speech, environmental sound, and music understanding tasks. 
For instance, Qwen-2.5-omni exhibits a substantial performance gap between its audio+image configuration and audio-only variants ($+15.55\%$, statistically significant across languages with paired t-test, $p < 0.01$), while the Gemma3n model shows slightly increased performance for Japanese, Malay, and Chinese. 
Sound category benefits the most from the visual information ($+11.71\%$ for Qwen-2.5-omni and $+4.31\%$ for Gemma3n among languages), which is expected as the sound-related questions often require more contextual information to be correctly answered. 
These findings suggest that visual grounding can provide complementary semantic cues for acoustic reasoning, especially for ambiguous audio events where contextual scene information disambiguates source identity, activity, or intent. 

\paragraph{Visual confusion occurs on cascaded modes.}
Despite the advantages of multimodal inputs, we identify a recurring failure mode in cascaded systems that process modalities independently before language reasoning. 
Cascaded variants of Gemma3n show degraded performance when image inputs are introduced alongside audio, indicating that visual information can distract rather than assist the model's reasoning process. 
This phenomenon is particularly visible in models adapted from speech-specific cases, where image conditioning sometimes lowers average accuracy across sound and music tasks. 
We hypothesize that fusion in cascaded pipelines induces representational competition: visual embeddings can be weakly correlated with the target acoustic semantics. 
Such visual confusion highlights the limitation of late-fusion designs and motivates researchers with tighter cross-modal alignment and shared latent representations.


\paragraph{Western languages perform better than eastern languages.}
A clear geographic and linguistic disparity emerges across the multilingual benchmark. 
Western languages, including German, English, and Spanish, generally obtain higher average scores than East Asian languages such as Japanese and Chinese across most model families. 
The trend is particularly pronounced for sound and music understanding subtasks, where performance degradation in Japanese and Chinese remains substantial even among frontier proprietary models. 
This discrepancy likely reflects uneven multilingual audio pretraining distributions, differences in phonetic structure, and the relative scarcity of culturally diverse non-speech acoustic supervision. 
Our findings suggest that multilingual multimodal competence remains strongly biased toward high-resource western language ecosystems.

\paragraph{Cross-Linguistic Evaluation.} 
We also observe a notable domain similarity structure across languages. 
Speech and sound performance display strong positive alignment (Pearson correlation $\approx 0.80$), implying that languages benefiting from robust spoken-audio understanding also tend to perform well on environmental sound tasks. 
Interestingly, Malay shows the strongest music understanding score ($62.65\%$) among all languages, suggesting that music-related reasoning differs from that in other domains. 
Overall, the results reveal that multilingual audio understanding is not uniformly distributed across domains. 
Speech generalization transfers relatively well across languages, whereas sound and music understanding expose residual multilingual gaps, particularly for eastern languages. 

\section{Ablation Study}
\label{sec:ablation}

\subsection{Ablation on Modality} 
\label{sec:ablation:modality}
First, we investigate the contribution of input modality by comparing multimodal (audio+image), audio-only, and image-only variants of Gemma3n-EXAM$^2$ on the EXAM$^2$-test set across six languages. 
Overall, multimodal conditioning consistently provides the strongest performance. 
Averaged across all languages, multimodal module achieves $61.22\%$, outperforming audio-only ($60.05\%$) and substantially exceeding image-only ($41.23\%$). 
The improvement is particularly pronounced for environmental sound understanding, where multimodal inputs yield systematic gains over audio-only inference in every language, indicating that visual grounding supplies complementary contextual cues for disambiguating non-speech acoustic events.

Interestingly, the relative benefit of modality differs across domains. 
Speech understanding achieves the best performance under audio-only setting, showing higher scores than audio+image in most languages. 
In contrast, multimodal fusion model provides clearer advantages for sound and, to a lesser extent, music understanding, where contextual scene information can help infer semantic intent, source identity, or acoustic structure. 
The non-trivial performance of image-only models suggests that visual priors encode weak but exploitable correlations between scenes and expected acoustic events. 
These findings reveal a complementary modality relationship: audio provides the primary semantic signal, while visual context improves robustness and cross-domain generalization, particularly for non-speech audio understanding.

\begin{table}
  \centering
  \resizebox{.48\textwidth}{!}{
  \begin{tabular}{lcccc}
    \toprule 
    \multirow{2}{*}{\textbf{Test Lang.}} & \multirow{2}{*}{\textbf{Train Lang.}} & \multirow{2}{*}{\textbf{Modality}} & \multicolumn{2}{c}{\textbf{Accuracy (\%)}} \\
    \cline{4-5}
    &&& Speech / Sound / Music & Avg.\\
    \midrule 
    \multirow{5}{*}{EN}&ALL& Audio+Image & \textbf{64.6 / 62.2 / 59.9} & \textbf{62.2} \\
    &EN & Audio+Image & 61.9 / 58.0 / 56.6 & 58.8\\
    &ALL & Audio & 67.0 / 57.7 / 61.1 & 61.9\\ 
    &ALL & Image & 40.5 / 43.5 / 43.4 & 42.5\\
    &-& Image & 30.3 / 33.9 / 38.5 & 34.3\\
    \midrule
    \multirow{5}{*}{DE}&ALL& Audio+Image & \textbf{66.4 / 62.8 / 59.9} & \textbf{63.0} \\
    &EN & Audio+Image & 48.9 / 49.2 / 45.8 & 47.8\\
    &ALL & Audio & 64.6 / 57.1 / 57.2 & 59.6\\ 
    &ALL& Image & 38.7 / 42.9 / 41.6 & 41.1\\
    &-& Image & 30.3 / 29.4 / 32.2 & 30.7\\
    \midrule
    \multirow{5}{*}{ES}&ALL& Audio+Image & \textbf{64.3 / 58.9 / 61.1} & \textbf{61.4} \\
    &EN & Audio+Image & 45.8 / 50.2 / 48.6 & 48.2\\
    &ALL & Audio & 68.2 / 57.1 / 57.5 & 60.9\\ 
    &ALL & Image & 40.5 / 45.3 / 41.9 & 42.6\\
    &-& Image & 35.1 / 29.4 / 36.8 & 33.8\\
    \midrule
    \multirow{5}{*}{JA}&ALL& Audio+Image & \textbf{63.4 / 56.8 / 59.6} & \textbf{59.9} \\
    &EN & Audio+Image & 49.5 / 43.8 / 43.1 & 45.5\\
    &ALL & Audio & 64.9 / 54.7 / 59.3 & 59.6\\ 
    &ALL & Image & 38.4 / 40.5 / 40.7 & 40.0\\
    &-& Image & 28.8 / 25.8 / 30.7 & 28.5\\
    \midrule
    \multirow{5}{*}{MS}&ALL& Audio+Image & \textbf{63.1 / 58.0 / 62.7} & \textbf{61.2} \\
    &EN & Audio+Image & 51.1 / 47.1 / 42.5 & 46.9\\
    &ALL & Audio & 65.8 / 55.9 / 58.1 & 59.9\\ 
    &ALL & Image & 41.7 / 45.0 / 40.1 & 42.3\\
    &-& Image & 27.9 / 27.3 / 30.4 & 28.6\\
    \midrule
    \multirow{5}{*}{ZH}&ALL& Audio+Image & \textbf{64.3 / 57.1 / 57.5} & \textbf{59.6}\\
    &EN & Audio+Image & 51.4 / 45.6 / 40.4 & 45.8\\
    &ALL & Audio & 67.0 / 53.8 / 54.5 & 58.4\\ 
    &ALL & Image & 38.1 / 40.2 / 38.3 & 38.9\\
    &-& Image & 29.2 / 32.1 / 35.2 & 31.9\\
    \bottomrule
  \end{tabular}
  }
  \caption{\label{ablation-modal}
    Ablation study for our Gemma3n-EXAM$^2$ on the EXAM$^2$-test by omitting different modalities on English-only and all six languages training data. 
    - denotes the model is evaluated via purely generated images without training on any language data. 
  }
\end{table}

\subsection{Ablation on Language}
\label{sec:ablation:language}
We further examine the effect of training language composition by comparing models trained on English-only (EN) data against models trained on all six languages (ALL) under identical multimodal settings. 
Multilingual training produces substantial and highly consistent improvements across all evaluation languages. 
Monolingual training leads to a degradation of $12.39$ points ($48.83\%$ averaged across languages). 
This pattern is universal across all six languages. 

Evidently, multilingual training benefits not only non-English languages but also English itself with a slight improvement of $+3.4$ points, suggesting multilingual exposure acts as a regularizer rather than introducing harmful language interference, improving representation learning even for the dominant training language.
A domain-wise breakdown further reveals that multilingual learning particularly strengthens sound and music understanding. 
For instance, in the German evaluation, multilingual training improves sound and music accuracies ($+13.6$ and $+14.1$ gains respectively). 
Similar gains appear across Japanese, Malay, and Chinese. 
We explain that multilingual supervision enriches the diversity of acoustic-language alignment during training, leading to representations that generalize better across culturally and linguistically heterogeneous auditory phenomena.


\subsection{Ablation on Purely Generated Images}
\label{sec:ablation:pure}
To investigate the contribution of human quality checking, we conduct an ablation study comparing the performance of our model evaluated on human-checked images versus purely generated images without the 25.05\% of human modified real screenshots. 
\Cref{ablation-modal} shows this solely image-generated setting leads to a substantial performance degradation across all six languages. 
In contrast, when the corresponding human-validated images are used, the average accuracies of the Gemma3n reach 46.70\% (EN), 44.79\% (DE), 45.69\% (ES), 44.79\% (JA), 45.29\% (MS), and 45.39\% (ZH), respectively. 
The results also show that training provides a clear benefit. 
Comparing the trained image-only setting (ALL) with the untrained setting (-), training improves average accuracy across all languages. 
This consistent improvement across languages indicates that multilingual training substantially strengthens the model's ability to extract and utilize information from the generated visual modality and suggests that manual visual refinement improves the quality of the visual modality rather than introducing exploitable generation artifacts. 



\section{Conclusion}
\label{sec:conclusion}
To summarize, we introduce EXAM$^2$, a comprehensive benchmark with in total $135,684$ multilingual and $22,614$ multimodal instances for evaluating audio understanding capabilities of \acp{LALM}, \acp{LLM}, \acp{MLLM}, and cascaded systems. 
We conduct extensive evaluations of open- and closed-source models across three audio domains (speech, sound, and music) and six languages (DE, EN, ES, JA, MS, and ZH). 
Results reveal that multimodal conditioning significantly enhances audio understanding, particularly for non-speech sound tasks, while multilingual training substantially improves performance across all languages, especially for eastern languages. 
Our proposed Gemma3n-EXAM$^2$ model, trained on our multilingual and multimodal training data, achieves competitive performance across all languages and audio categories, demonstrating the high quality of our benchmark for improving multilingual and multimodal audio understanding. 
Future work can explore more efficient training methods and a wider range of languages to further enhance the multilingual and multimodal capabilities for the EXAM$^2$ benchmark.

\section*{Limitations}
\label{sec:limitation}
There are several limitations to our current evaluation of the EXAM$^2$ benchmarks. 
In these benchmarks, we have prioritized accuracy over efficiency. 
Future evaluations should incorporate inference speed and deployment constraints to enable a more comprehensive assessment of model performance in real-world applications. 
Second, due to limited computational resources, we have only curated a subset of the AudioMCQ dataset to generate visual representations for training our model. 
Future work can explore the full dataset and consider more efficient training methods to leverage the entire dataset effectively. 
Additionally, our language coverage is limited to European and Asian languages, a wider range of languages such as African language families can be included to better assess the multilingual capabilities of models on the EXAM$^2$ benchmarks. 

\section*{Ethics Statement}
\label{sec:ethic}
The authors are not purposely creating or using any visual data that contains personally identifiable information or offensive content. 
Generated images are solely used for training and evaluation purposes on the EXAM$^2$ benchmarks, and we have taken care to ensure that they do not contain any harmful or inappropriate content.

\section*{Acknowledgements}
\label{sec:acknowledgement}
This research is supported by UDF01004676 and the National Research Foundation, Singapore under its National Large Language Models Funding Initiative (AISG Award No: AISG-NMLP-2024-004), and by A*STAR under its RIE2025 Japan-Singapore Joint Call: Japan Science and Technology Agency (JST) and Agency for Science, Technology and Research (A*STAR) 2024 (Award R24I6IR136). Any opinions, findings and conclusions or recommendations expressed in this material are those of the author(s) and do not reflect the views of National Research Foundation, Singapore. 

\section*{Use of AI Tools}
\label{sec:ai}
We have used GPT-5-mini for translating the original English questions and choices into other five languages, and we have used GPT-image-2 for generating the visual representations of the audio. 
The authors also acknowledge the use of ChatGPT for assistance with grammar, punctuation, and vocabulary refinement, as well as debugging tasks. 

\bibliography{custom}

\clearpage
\appendix

\section{Appendix}
\label{sec:appendix}

\subsection{Dataset Details}
\label{sec:appendix:data}

\subsubsection{Translation Guidelines}
\label{sec:appendix:data:translation}
When translating the original English questions and choices into other five languages, we first use the GPT-5-mini to follow the instruction in \Cref{fig:p1}. 
The translated results are then carefully reviewed by native speakers of the target languages (or researchers who are fluent in those languages for over 5 years) to ensure the accuracy, fluency, and appropriateness of the translations.
Any issues identified during the review process were addressed and corrected to ensure the highest quality of the translated questions and answer choices. 

Below is our specific guideline to ensure the quality and consistency of the translations: 

\begin{enumerate}
  \item \textbf{Preserve Meaning}: The translation should accurately convey the meaning of the original English text, including the question and all answer choices. 
  \item \textbf{Maintain Format}: The structure of the question and answer choices should be preserved in the translation, including the numbering and formatting. 
  \item \textbf{Use Natural Language}: The translation should use natural and fluent language that is appropriate for the target language, avoiding literal translations that may sound awkward or unnatural.
  \item \textbf{Cultural Relevance}: If the original question contains culturally specific references or examples, we may adapt them to be more relevant to the target language and culture while maintaining the original intent of the question. 
  \item \textbf{Consistency Across Languages}: We aimed to maintain consistency in the translation style and terminology across all languages to ensure that the questions are comparable and that the evaluation is fair across different linguistic contexts. 
\end{enumerate}

\subsection{Model Overview}
\label{sec:appendix:model}

\subsubsection{Multilingual Examples}
\label{sec:appendix:data:lang}

\Cref{multilingual} and \Cref{transcap} shows the multilingual translation validated by native speakers on the EXAM$^2$-test set. 
By not only checking the correctness and fluency of the translations, we also ensure that the translated answer in a consistent position with the original English answer. 
For instance, we manually removed the redundant indexes of the given options, especially for the German and Spanish, as GPT-5-mini tends to automatically add the inconsistent indexes in the translated choices, which may cause confusion for the model. 
Additionally, if the audio contains some specific words that are asked in the question, we keep those words unchanged in the translated questions and choices, as shown in \Cref{transcap}. 
This step aligns with our processed transcript and caption, as most of the audio contains English words, and we want to ensure the consistency between the content within languages.

\begin{figure}
  \centering
   \begin{lightbox}
  \begin{footnotesize} 
  \itshape   
Translate the following audio understanding/reasoning question, its choices, and its answer into \{Language\}.\\
Keep technical/music notation unchanged. \\
Keep the English words in the choices unchanged if they are asked which word appears first in the question or something similar.\\
Make sure the translated answer matches one of the translated choices.

\medskip

Question:\\
\{question\}

\medskip

Choices:\\
\{option 1. \}
\{option 2. \}
...

\medskip

Answer:\\
\{answer\}

  \end{footnotesize}
\end{lightbox}
   \caption{Translation prompt of GPT-5-mini. \{Language\} is replaced by the target language (German, Spanish, Japanese, Malay, and Chinese) when applying the prompt.} 
   \label{fig:p1}
\end{figure}

\begin{figure}
  \centering
   \begin{lightbox}
  \begin{footnotesize} 
  \itshape   
A real-world and precise visual representation of the audio: \{option 1. \}, \{option 2. \}, ....
  \end{footnotesize}
\end{lightbox}
   \caption{Image generation prompt of GPT-image-2 and stable diffusion.} 
   \label{fig:p2}
\end{figure} 

\begin{figure}[ht]
  \centering
   \begin{lightbox}
  \begin{footnotesize} 
  \itshape   
You are an expert in multimodal audio understanding. Let us solve a multiple-choice question together.

Image 1 -> Choice 1 \\
Image 2 -> Choice 2 \\
Image 3 -> Choice 3 \\
Image 4 -> Choice 4 \\

Use ALL inputs:\\
- Question \\
- Choices \\
- Transcript \\
- Caption \\
- Images \\

Instructions:\\
- Select the SINGLE best answer. \\
- Return ONLY the exact choice text. \\
- Do not output the choice number. \\
- Do not explain. \\
- Do not output anything else. \\

Question:\\
\{question\}

\medskip

Choices:\\
\{choices\_text\}

\medskip

Transcript:\\ 
\{transcript\}

\medskip

Caption:\\
\{caption\}
  \end{footnotesize}
\end{lightbox}
   \caption{Inference prompt for multimodal models including Phi4-multimodal-instruct, Qwen2.5-omni, and Gemma3n.} 
   \label{fig:p3}
\end{figure} 

\begin{figure}[ht]
  \includegraphics[width=\linewidth]{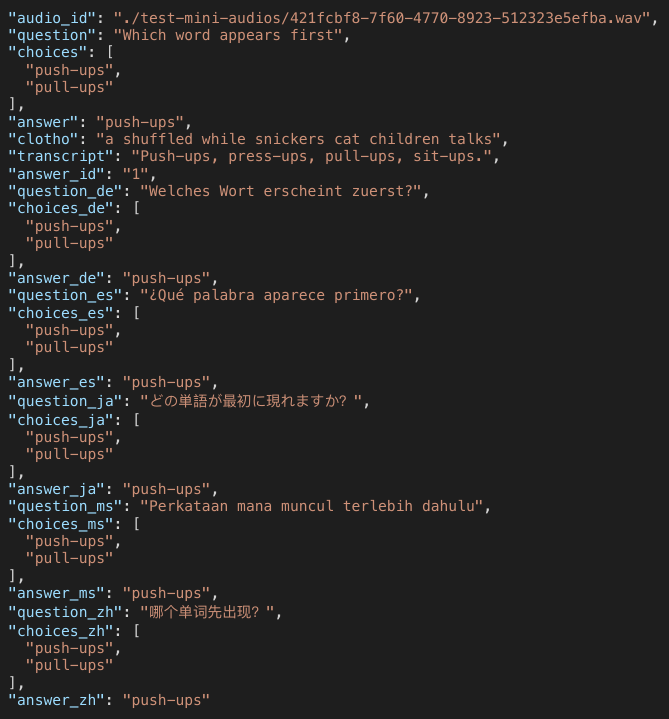}
  \caption{Transcript and caption, multilingual translation examples in \texttt{json} screenshot on our EXAM$^2$.}
  \label{transcap}
\end{figure}

\begin{figure*}[ht]
  \includegraphics[width=\linewidth]{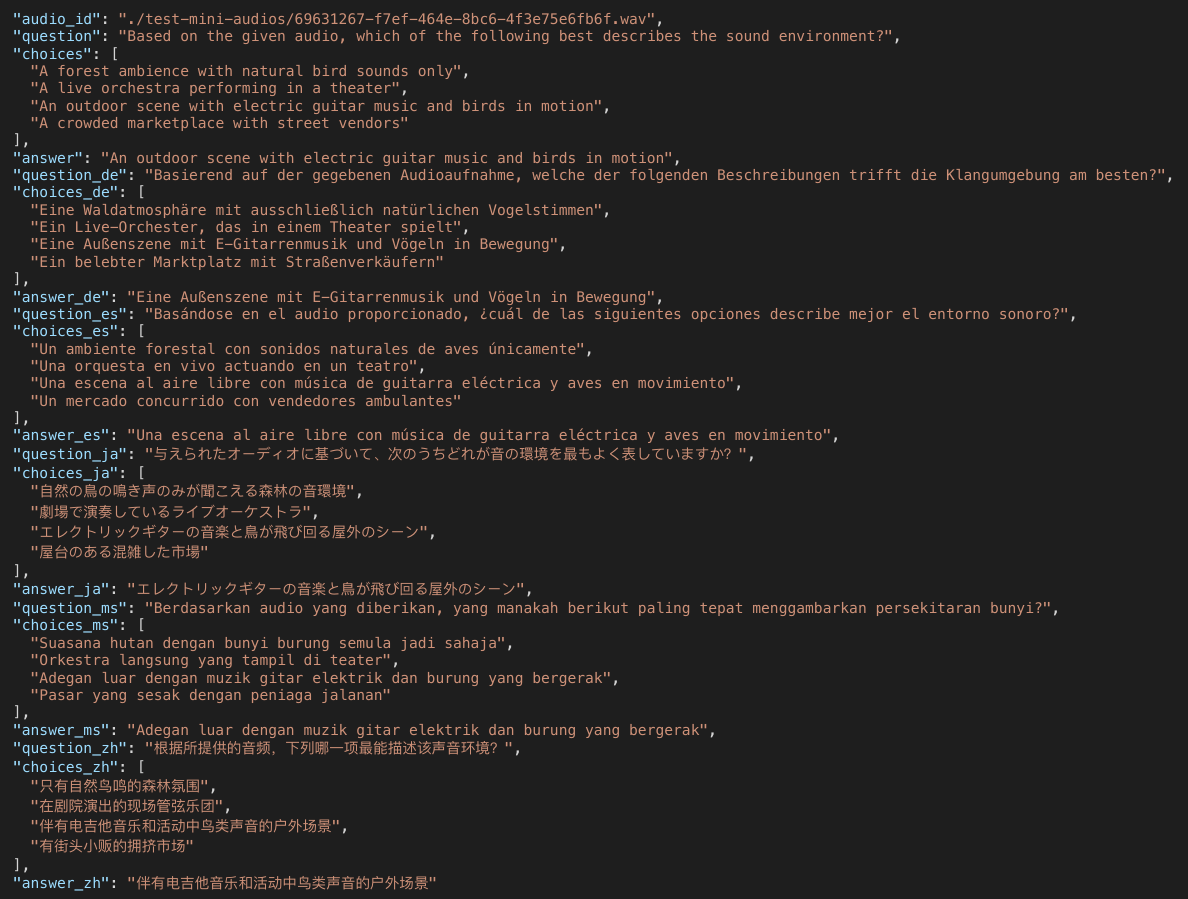}
  \caption{Multilingual translation examples in \texttt{json} screenshot on our EXAM$^2$.}
  \label{multilingual}
\end{figure*}

\subsubsection{Visual Examples}
\label{sec:appendix:data:visual}

\begin{figure*}[t]
  \centering
  \begin{subfigure}[b]{\linewidth}
    \centering
    \includegraphics[width=.24\linewidth]{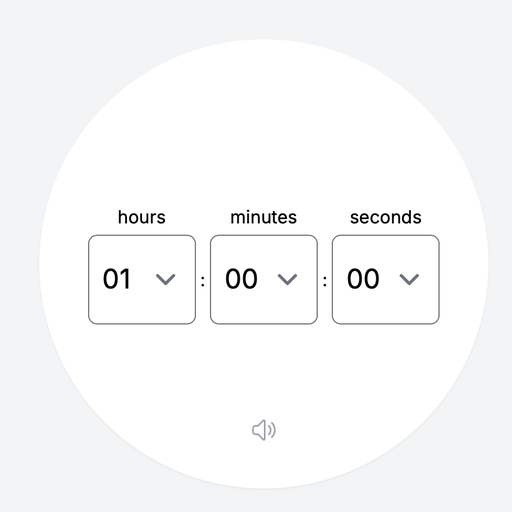}
    \includegraphics[width=.24\linewidth]{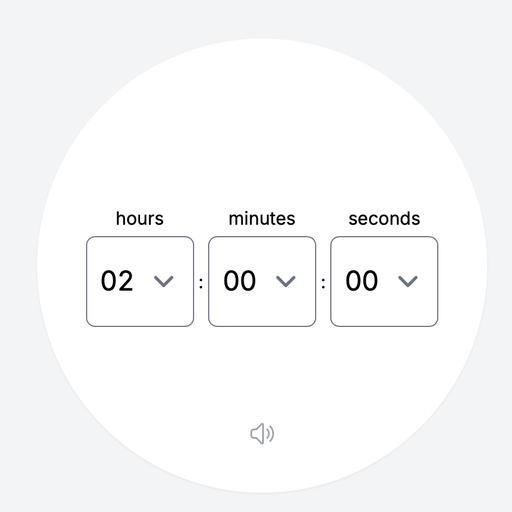}
    \includegraphics[width=.24\linewidth]{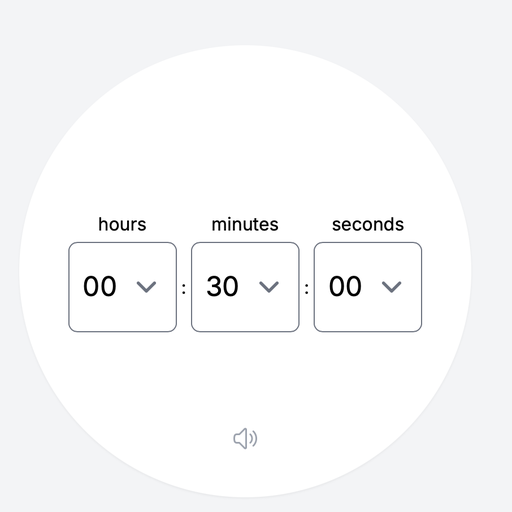}
    \includegraphics[width=.24\linewidth]{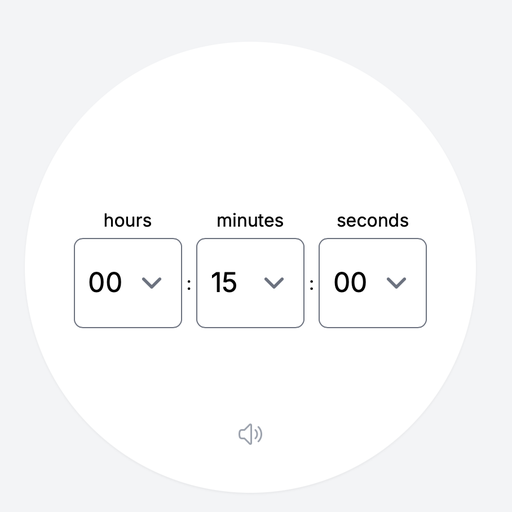}
    \caption{Visual examples of speech category on EXAM$^2$ evaluation dataset. 
    The generated images are from choices prompts (from left to right): 
    \texttt{An hour; Two hours; Thirty minutes; Fifteen minutes.}}
    \label{mmau0}
  \end{subfigure}
  \hfill
  \begin{subfigure}[b]{\linewidth}
    \centering
    \includegraphics[width=.24\linewidth]{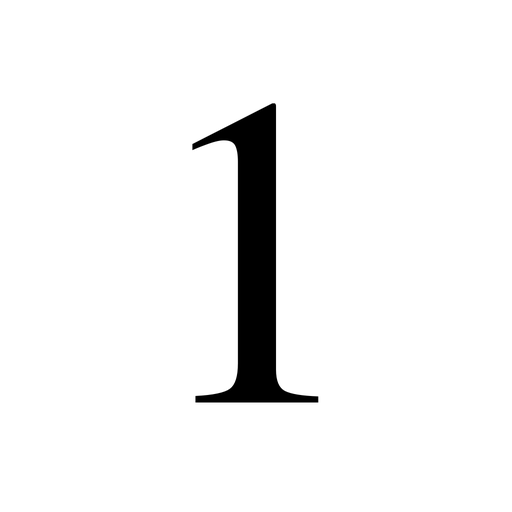}
    \includegraphics[width=.24\linewidth]{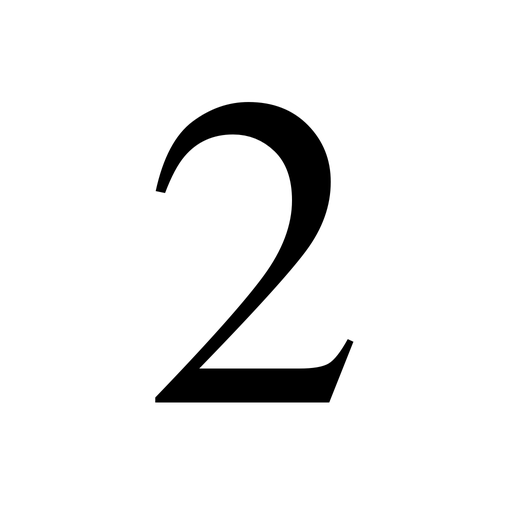}
    \includegraphics[width=.24\linewidth]{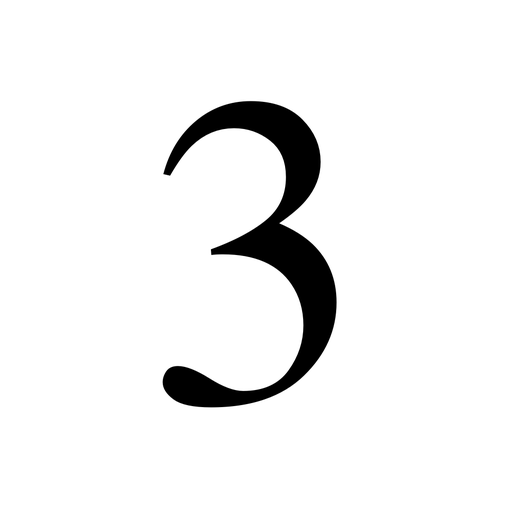}
    \includegraphics[width=.24\linewidth]{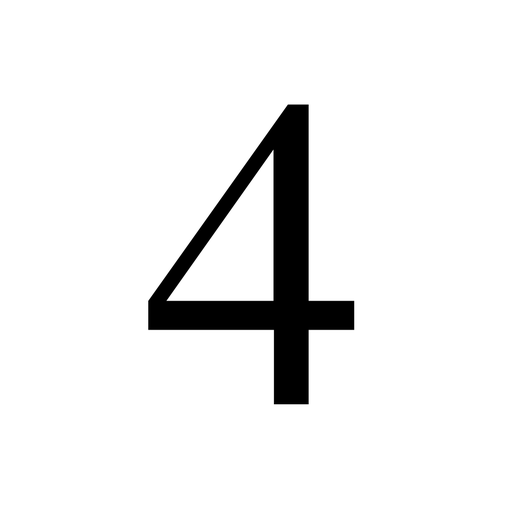}
    \caption{Visual examples of sound category on EXAM$^2$ evaluation dataset. 
    The generated images are from choices prompts (from left to right): 
    \texttt{Once; Twice; Three times; Four times.}}
    \label{mmau1}
  \end{subfigure}
  \hfill
  \begin{subfigure}[b]{\linewidth}
    \centering
    \includegraphics[width=.24\linewidth]{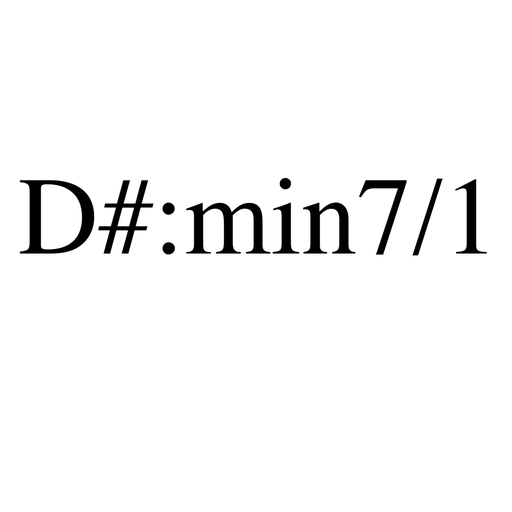}
    \includegraphics[width=.24\linewidth]{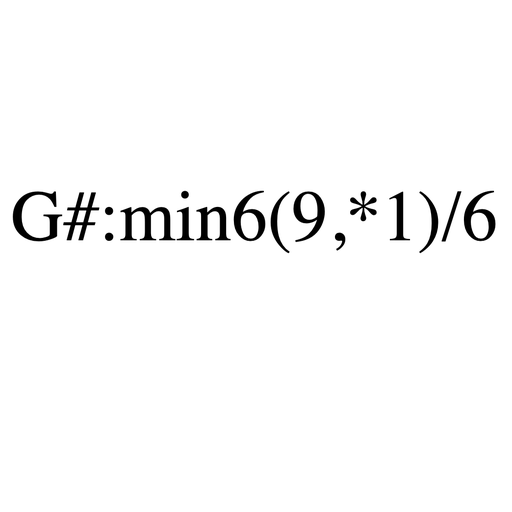}
    \includegraphics[width=.24\linewidth]{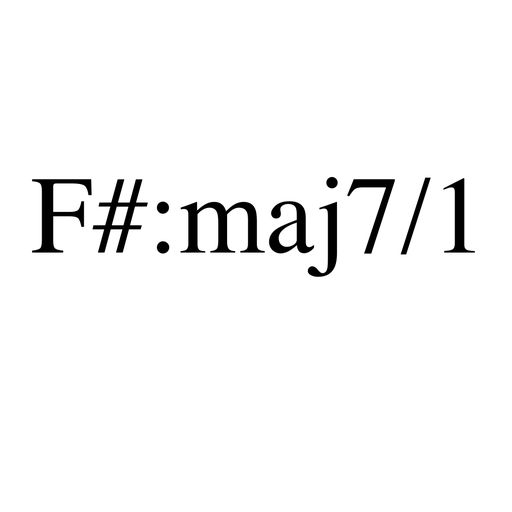}
    \includegraphics[width=.24\linewidth]{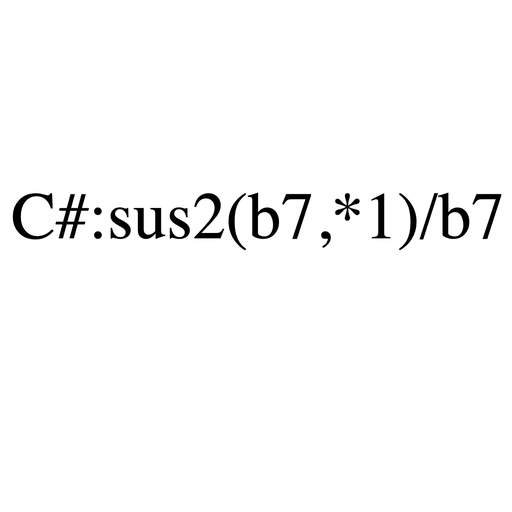}
    \caption{Visual examples of music category on EXAM$^2$ evaluation dataset. 
    The generated images are from choices prompts (from left to right): 
    \texttt{D\#:min7/1; G\#:min6(9,*1)/6; F\#:maj7/1; C\#:sus2(b7,*1)/b7.}}
    \label{mmau2}
  \end{subfigure}
  \caption {Visual examples manually created on EXAM$^2$ evaluation set.}
  \label{mmau}
\end{figure*}

\begin{figure*}[t]
  \centering
  \begin{subfigure}[b]{\linewidth}
    \centering
    \includegraphics[width=.24\linewidth]{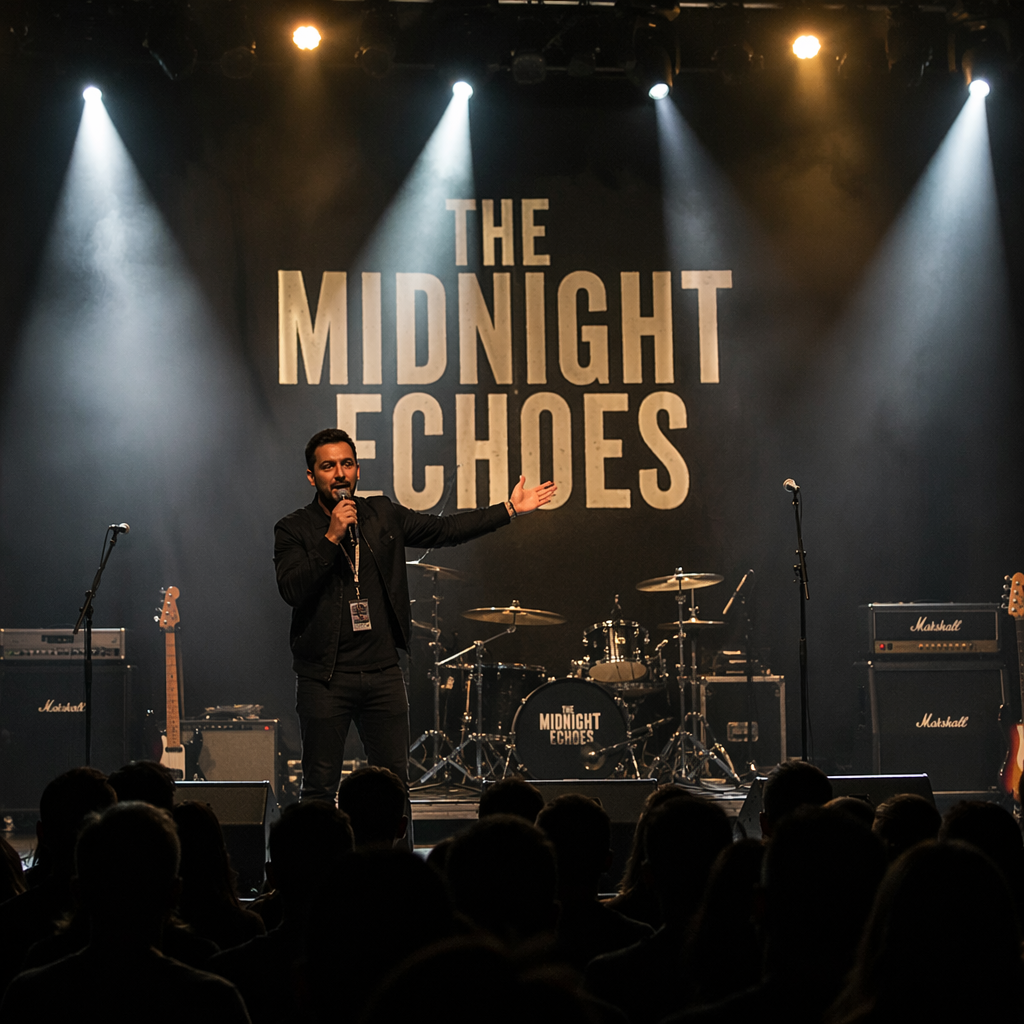}
    \includegraphics[width=.24\linewidth]{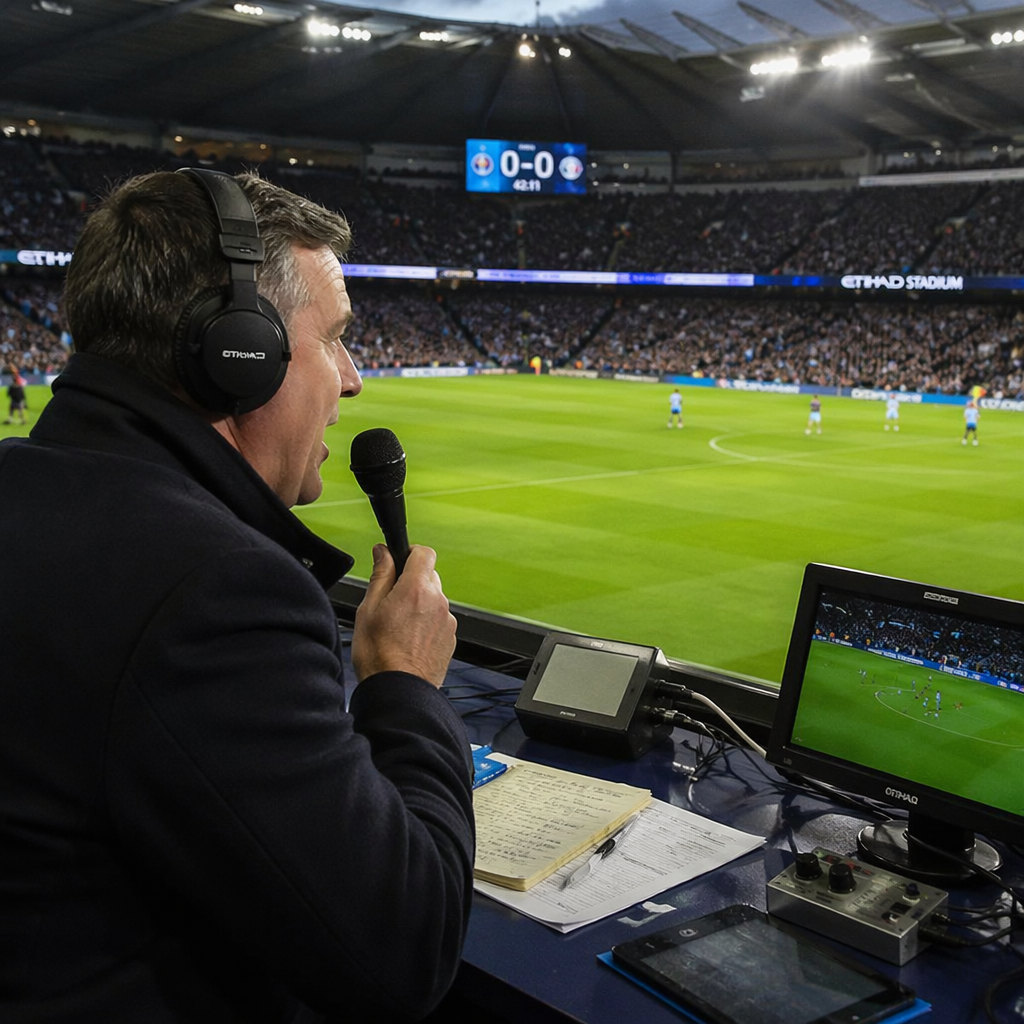}
    \includegraphics[width=.24\linewidth]{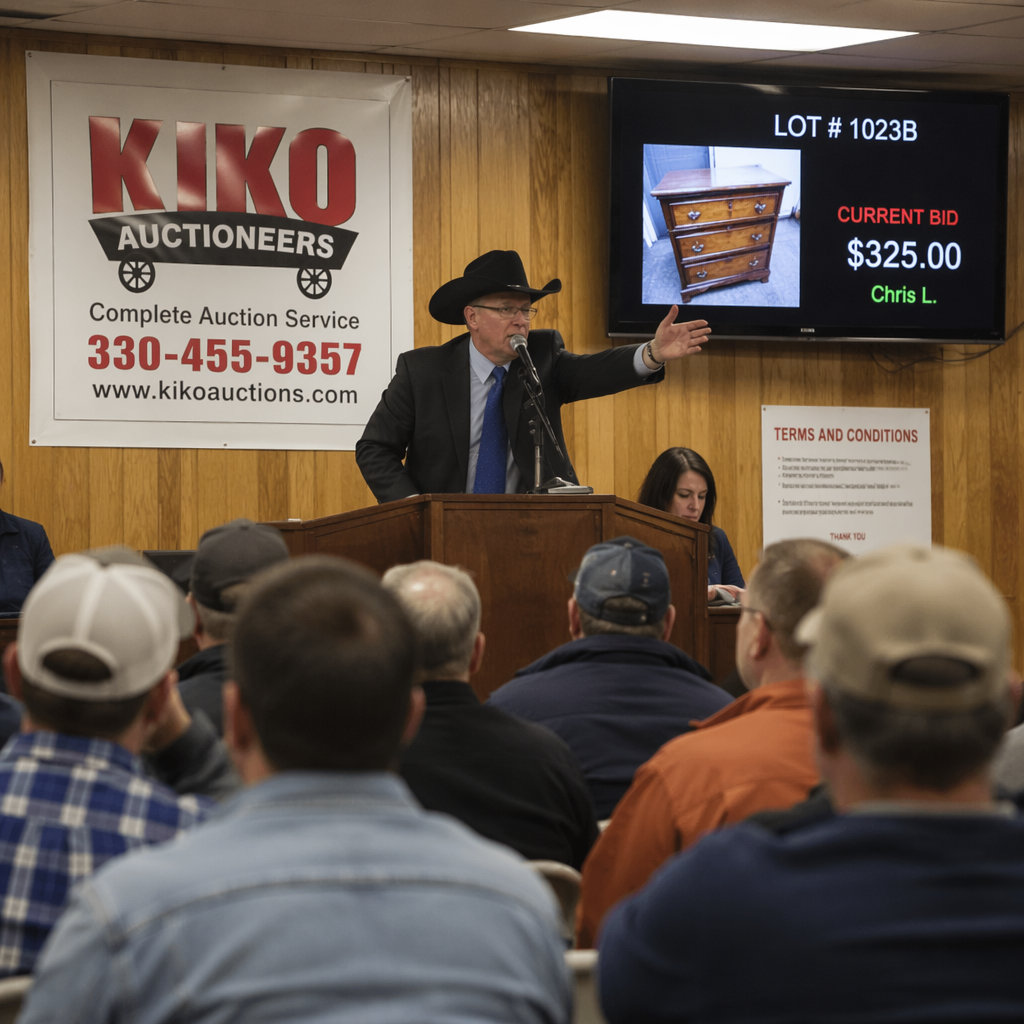}
    \includegraphics[width=.24\linewidth]{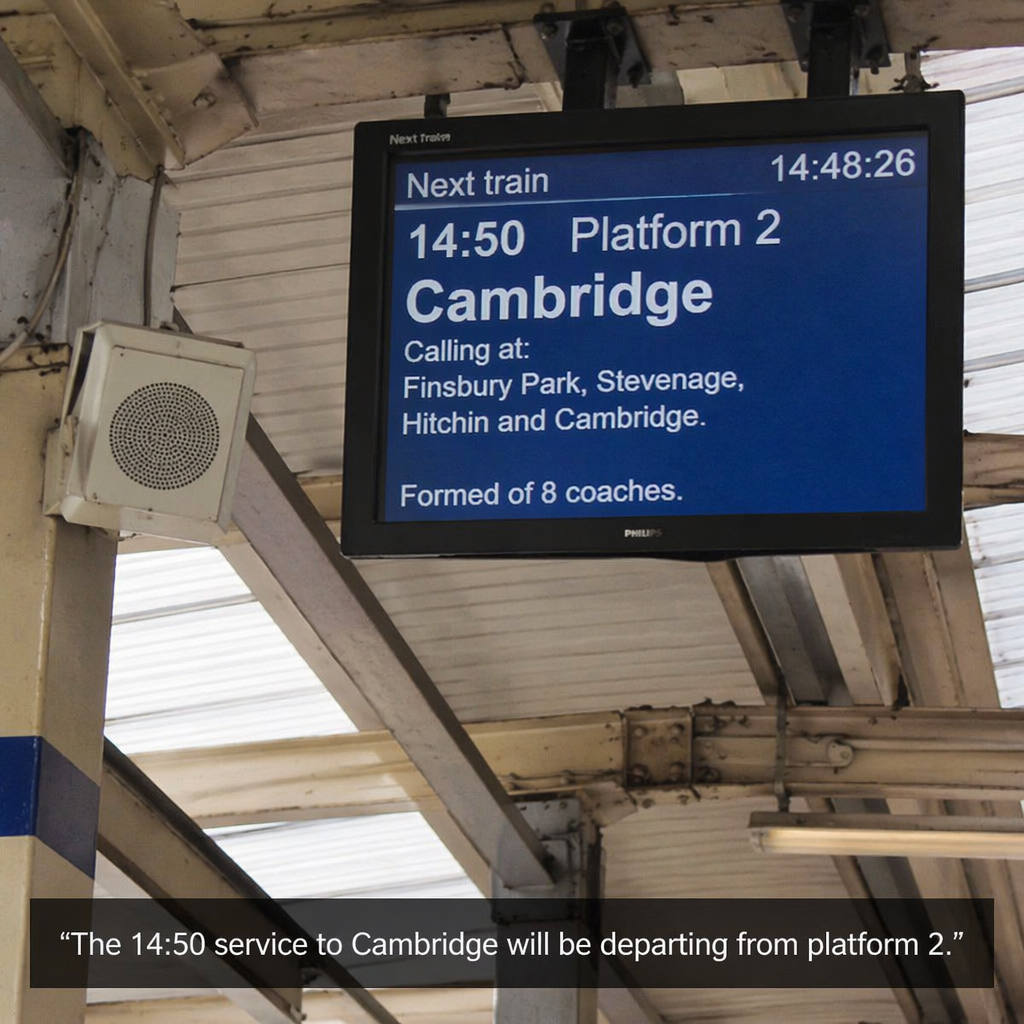}
    \caption{Visual examples of speech category on EXAM$^2$-Clotho training dataset. 
    The generated images are from choices prompts (from left to right): 
    \texttt{A concert host introducing a band; A sports commentator describing a match; An auction announcer addressing a crowd; A train station departure announcer.}}
    \label{eg0}
  \end{subfigure}
  \hfill
  \begin{subfigure}[b]{\linewidth}
    \centering
    \includegraphics[width=.24\linewidth]{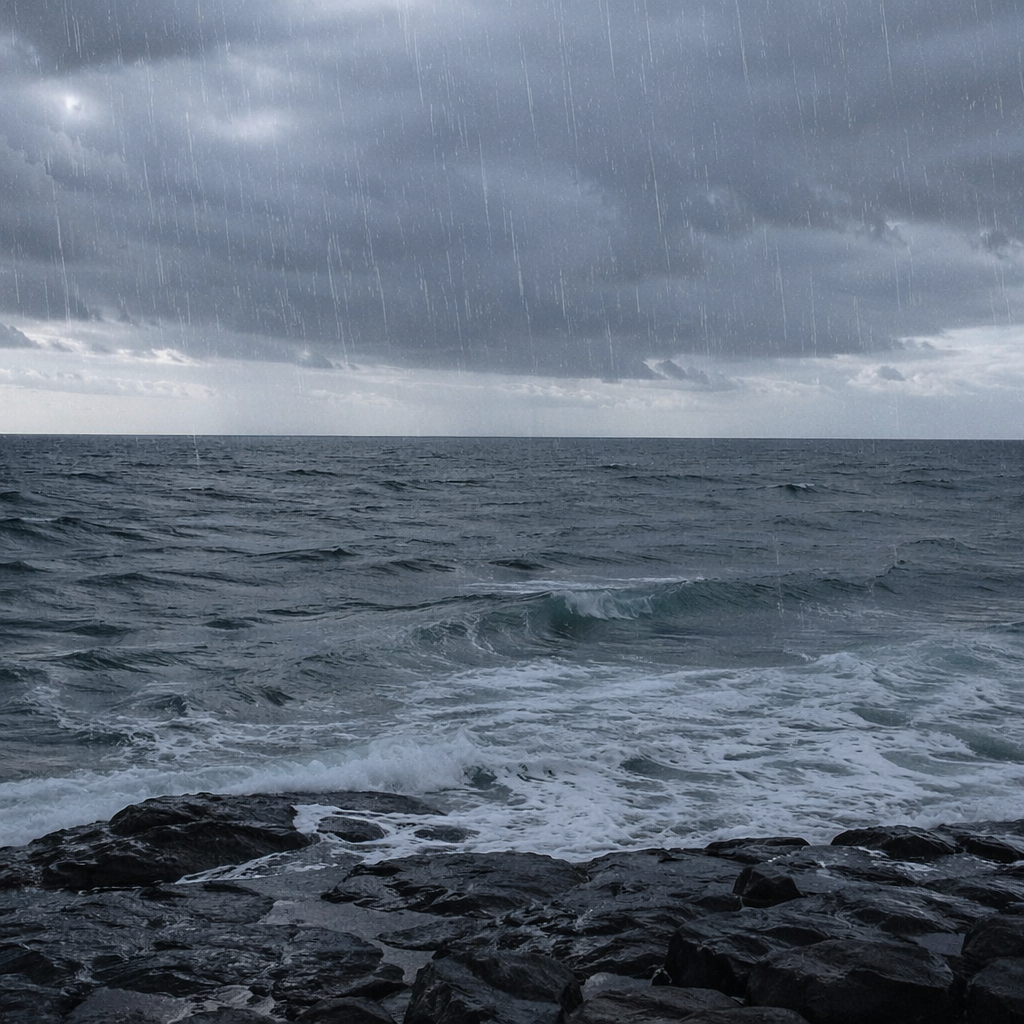}
    \includegraphics[width=.24\linewidth]{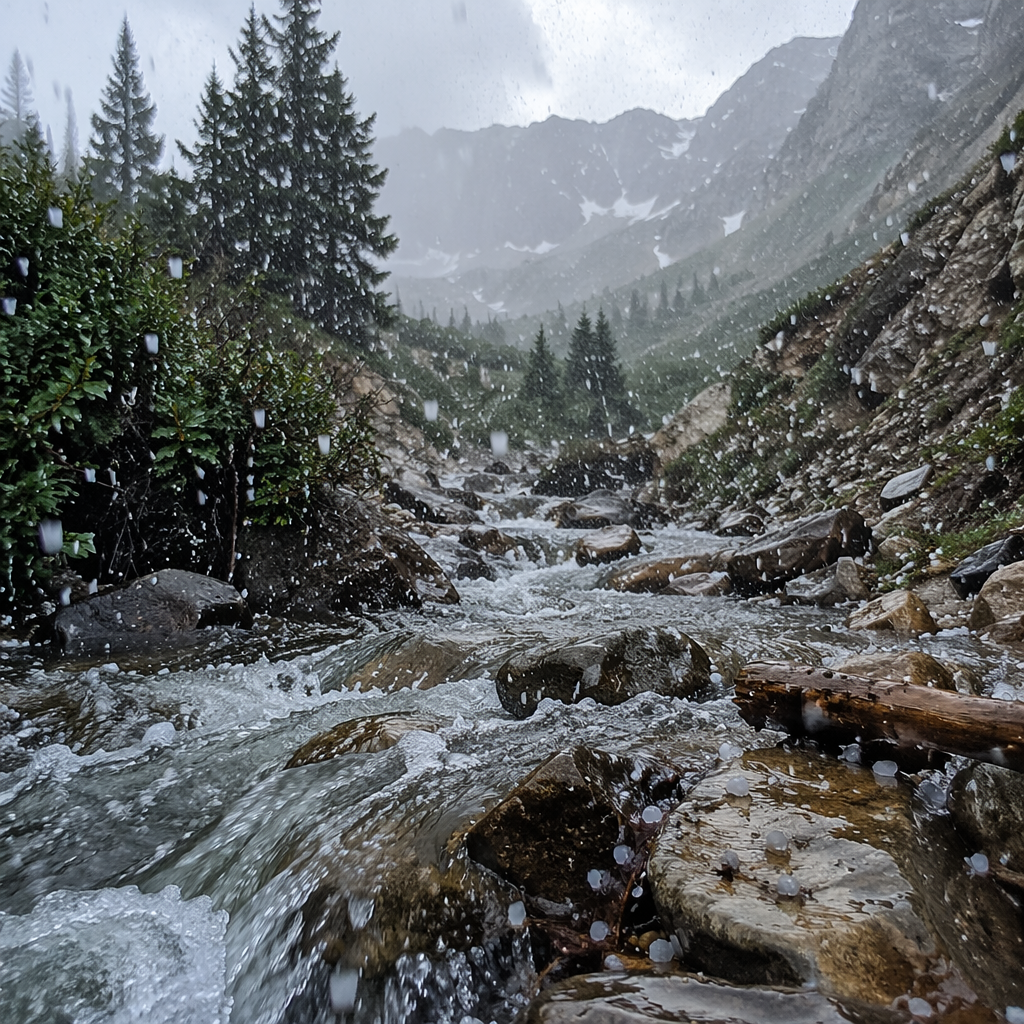}
    \includegraphics[width=.24\linewidth]{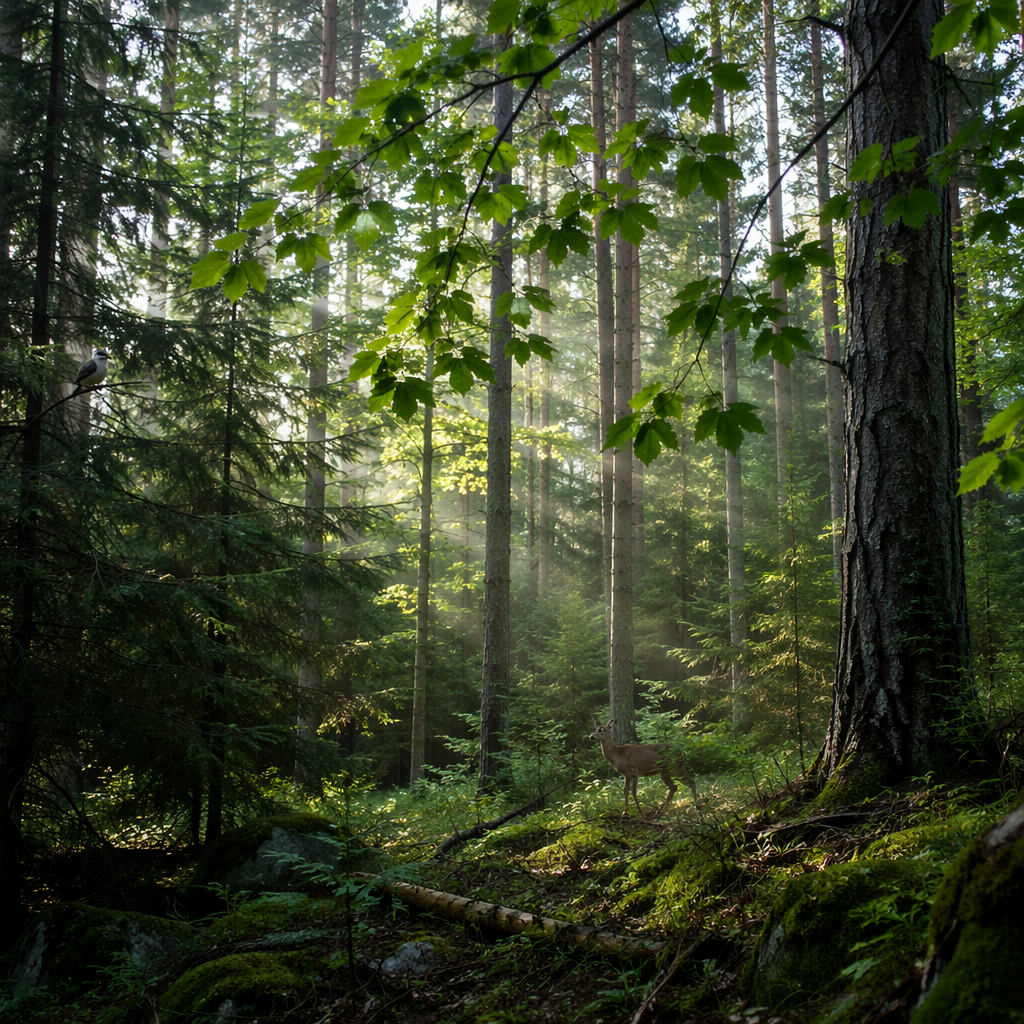}
    \includegraphics[width=.24\linewidth]{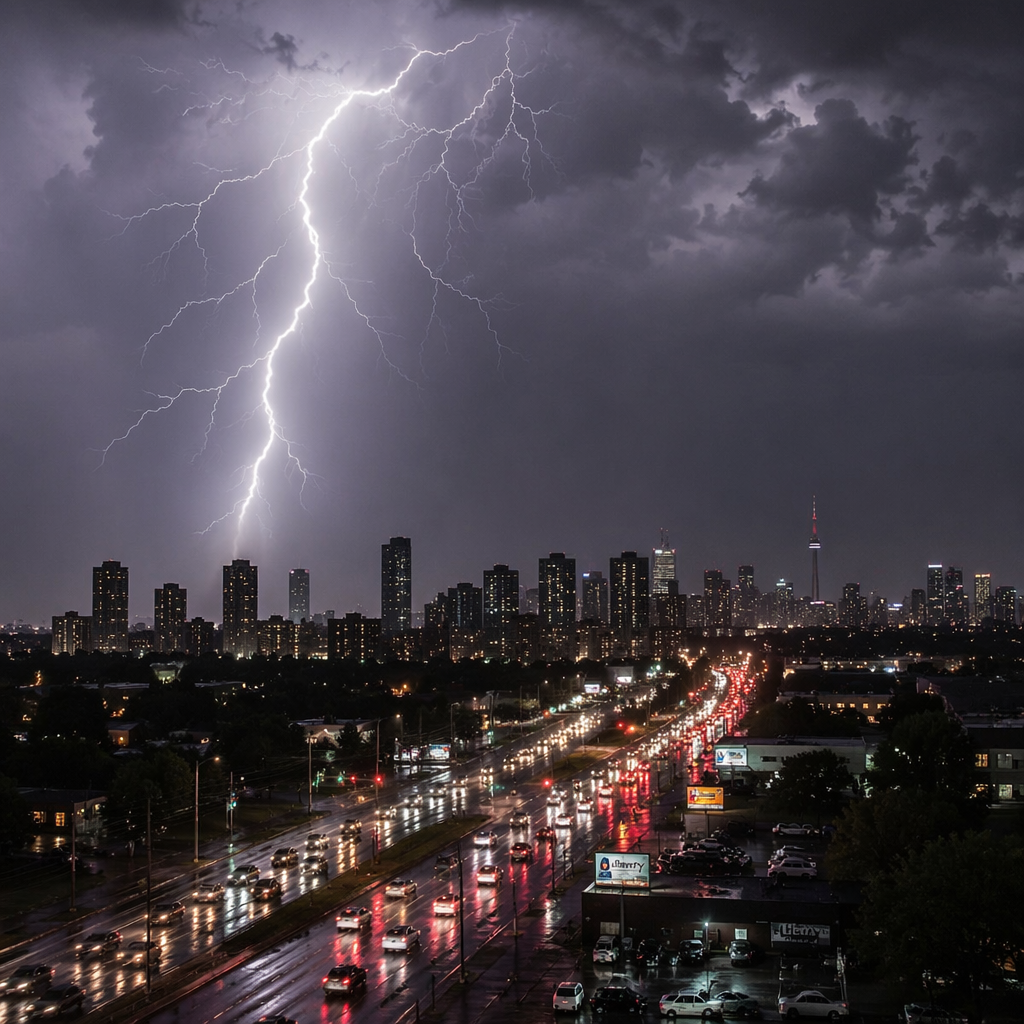}
    \caption{Visual examples of sound category on EXAM$^2$-Clotho training dataset. 
    The generated images are from choices prompts (from left to right): 
    \texttt{The rain and ocean waves; The hail and mountain stream; The wind and forest animals; The thunder and city traffic.}}
    \label{eg1}
  \end{subfigure}
  \hfill
  \begin{subfigure}[b]{\linewidth}
    \centering
    \includegraphics[width=.24\linewidth]{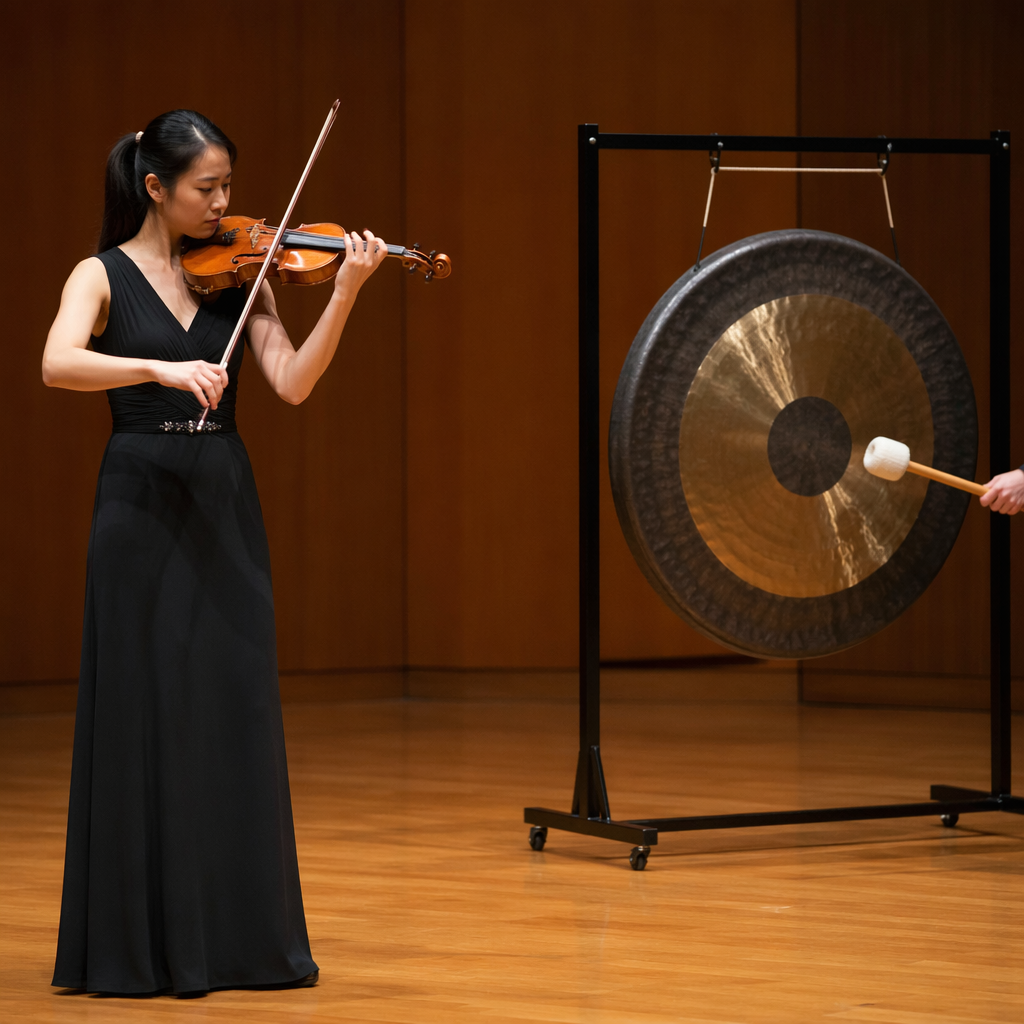}
    \includegraphics[width=.24\linewidth]{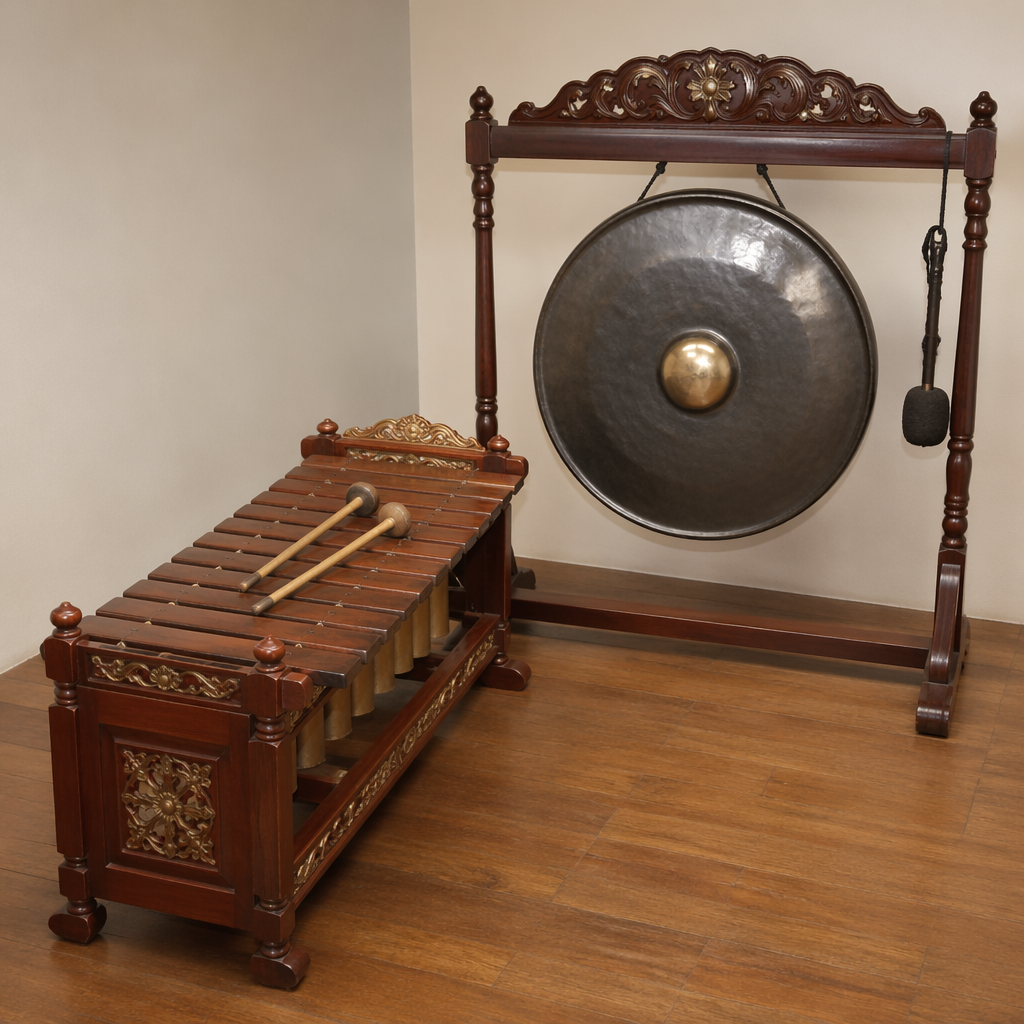}
    \includegraphics[width=.24\linewidth]{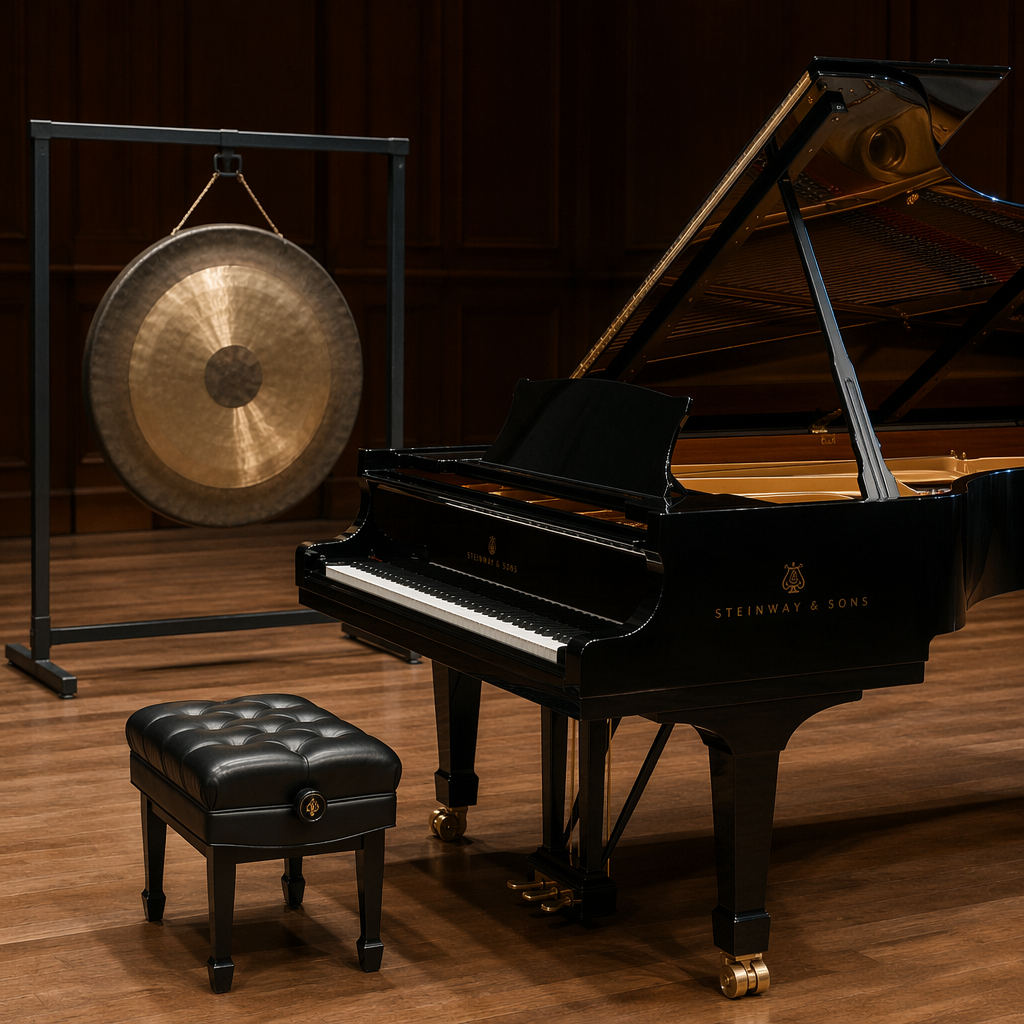}
    \includegraphics[width=.24\linewidth]{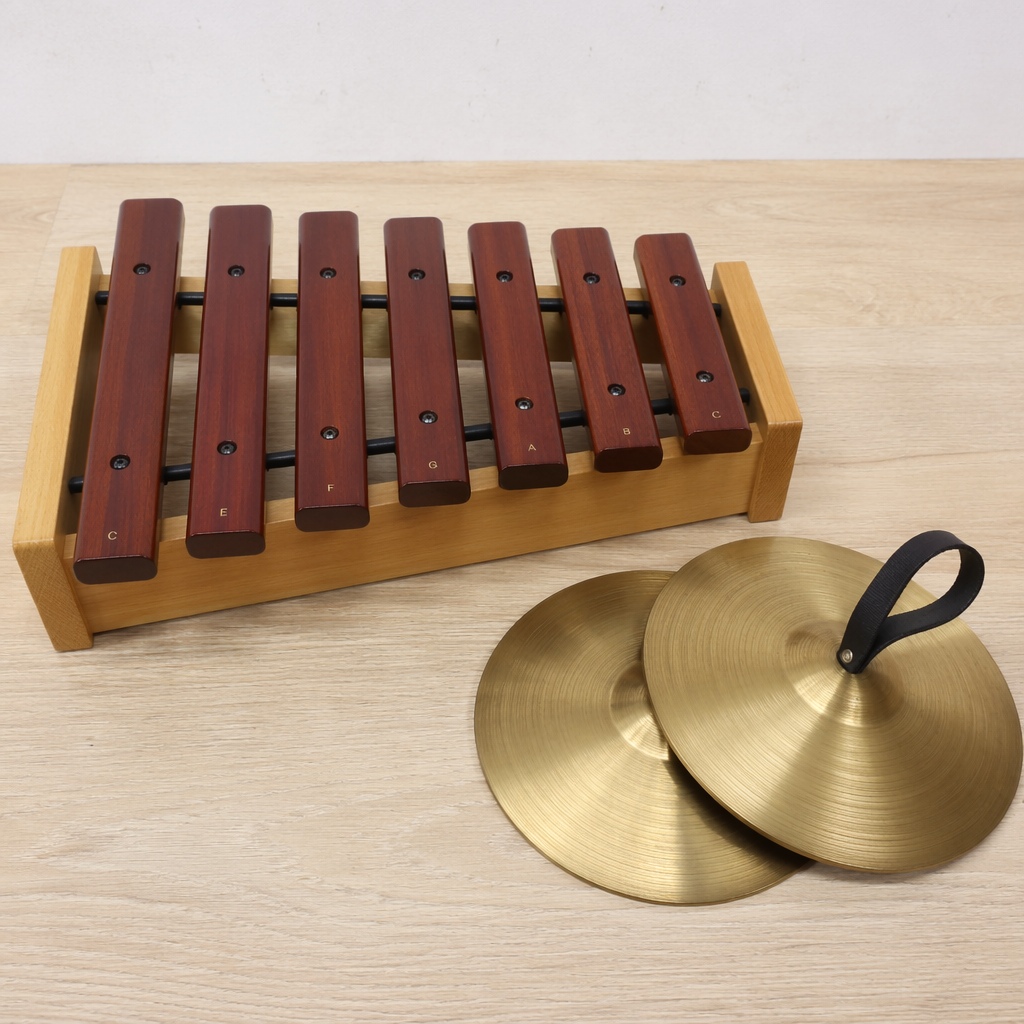}
    \caption{Visual examples of music category on EXAM$^2$-Clotho training dataset. 
    The generated images are from choices prompts (from left to right): 
    \texttt{The violin and the gong; The xylophone and the gong; The piano and the gong; The xylophone and the cymbals.}}
    \label{eg2}
  \end{subfigure}
  \caption {Visual examples generated by GPT-image-2 on EXAM$^2$-Clotho training set.}
  \label{clotho}
\end{figure*}

We provide visual examples of generated images for all three audio categories (speech, sound, and music) in \Cref{mmau} and \Cref{clotho}. 
For images in \Cref{mmau0}, \Cref{mmau1}, and \Cref{mmau2}, they are screenshots manually created by authors, as the original images generated by stable diffusion/GPT-image-2 are not highly relevant to the audio content. 
However, due to strong capability of visual understanding in environmental sounds, GPT-image-2 can generate more fine-grained images for sound understanding questions, which is particularly helpful in our EXAM$^2$-train, as shown in \Cref{clotho}.

\subsection{Prompts}
\label{sec:appendix:prompt}

\Cref{fig:p1} shows the prompt we used for translating the original English questions and choices into other five languages (DE, ES, JA, MS, and ZH) using GPT-5-mini.
For simplicity, prompt to generate our image using GPT-image-2 and stable diffusion is shown in \Cref{fig:p2}. 

When evaluating the performance of multimodal models, we use the same prompt as shown in \Cref{fig:p3} for all multimodal models. 
Specifc modalities (e.g., image-only or audio-only) are controlled by removing the corresponding input from the prompt. 
For instance, for image-only evaluation, we only keep the question, choices, and images in the prompt, while for audio-only evaluation, we only keep the question, choices, and original audio in the prompt. 
Cascaded evaluation is conducted by applying the transcript / caption instead of the audio input in the prompt. 

\subsection{Experiment Details}
\label{sec:appendix:experiment}


\begin{table}
  \centering
  \begin{tabular}{lr}
    \toprule 
    \textbf{Hyperparameter}           & \textbf{Value} \\
    \midrule 
      Data Type & bfloat16 \\
      Learning Rate & 2e-4 \\
      Batch Size & 4 \\
      Gradient Accumulation Steps & 8 \\
      Training Epochs & 6 \\
      Lora Alpha & 32 \\
      Lora Dropout & 0.05 \\
    \bottomrule
  \end{tabular}
  \caption{\label{hyper}
    Hyperparameters used for supervised fine-tuning of Gemma3n.
  }
\end{table}

We give additional detailed hyperparameters used in our experiments in \Cref{hyper}. 
The backbone model for fine-tuning is Gemma3n-E4B, and we use the OmniLoRA method for efficient fine-tuning.

\end{document}